\pdfoutput=1
\documentclass[12pt,a4]{article}
\usepackage{latexsym,epsfig,amssymb,euscript,amsmath,verbatim,cite}

\newcommand{\be}{\begin{equation}}
\newcommand{\ee}{\end{equation}}
\newcommand{\bea}{\begin{eqnarray}}
\newcommand{\eea}{\end{eqnarray}}

\numberwithin{equation}{section}
\usepackage{bm}
\usepackage{hyperref}
\usepackage{color}

\begin{document}
\pagestyle{empty}

\vspace{1.8cm}

\begin{center}
{\LARGE{\bf Electromagnetic response \\\vspace{3pt} of strongly coupled plasmas}}

\vspace{1cm}

{\large{Davide Forcella$^{a,b,}$\footnote{\tt  dforcell@ulb.ac.be},
Andrea Mezzalira$^{a,}$\footnote{\tt andrea.mezzalira@ulb.ac.be}, 
Daniele Musso$^{c,}$\footnote{\tt dmusso@ictp.it} 
\\[1cm]}}

{\small{{}$^a$  Physique Th\'eorique et Math\'ematique\\
Universit\'e Libre de Bruxelles, C.P. 231, 1050 Bruxelles, Belgium}\\
\medskip
\small{{}$^b$ Laboratoire de Physique Th\'eorique et Hautes Energies, Universit\'e Pierre et Marie Curie,\\
4 Place Jussieu, 75252 Paris Cedex 05, France\\}
\medskip
\small{{}$^c$ Abdus Salam International Centre for Theoretical Physics (ICTP)\\
        Strada Costiera 11, I 34014 Trieste, ITALY\\}}

\vspace{1cm}

{\bf Abstract}
\end{center}

We present a thorough analysis of the electromagnetic response of strongly coupled neutral plasmas described by the gauge/gravity correspondence. 
The coupling of the external electromagnetic field with the tower of quasi-normal modes of the plasmas supports the presence of various 
electromagnetic modes with different properties. Among them we underline the existence of negative refraction with 
low dissipation for a transverse non-hydrodynamical mode. Previous hydrodynamical approaches have shown the ubiquitous 
character of negative refraction in charged plasmas and the absence thereof in neutral plasmas. Our results here extend
the analysis for neutral plasmas beyond the hydrodynamical regime. 
As an application of these new insights we briefly discuss the case of the quark gluon plasma in the temperature dominated regime.

\noindent

\newpage

\setcounter{page}{1} \pagestyle{plain} \renewcommand{\thefootnote}{\arabic{footnote}} \setcounter{footnote}{0}

\tableofcontents

\section{Introduction}

The study of the linear response of a system to a small external perturbation 
is an essential tool to gain insight on the collective behavior of its constituents 
and to provide information about the transport properties at thermal equilibrium%
\footnote{The topic of linear response is of course extremely wide, 
for the fundamental material of use here we refer to standard text-books 
and in particular to \cite{FosterKM}.}.
Media having electrically movable components that can easily respond 
to an external electromagnetic field are ubiquitous in physics: high-energy physics plasmas and condensed matter 
superconductors are instances of this sort.
For such media usually the electromagnetic field interacts with thermal and mechanical modes; 
hence the electromagnetic energy finds various channels of propagation and dissipation through the medium. 
The response of the system to an external electromagnetic field is then in general non-local 
and the medium is said to be spatially dispersive (see for instance \cite{Landau}). 
This rich dynamics generates an intriguing phenomenology which can feature exotic electromagnetic effects such as negative refraction (NR) \cite{SpatNeg} and additional light-waves (ALW) \cite{ALW} as we will shortly review.

Exotic electromagnetic phenomena have recently attracted an intense theoretical and experimental 
(and even technological) interest especially in relation to artificial systems called meta-materials \cite{Metamat}.
Inspired by the amazing developments obtained on meta-materials, in \cite{Noi1,Noi2, Noi3} it was first 
discovered that charged fluids which admit a hydrodynamical description present exotic electromagnetic properties
as a general feature. The negative refraction phenomenon (in a certain range of frequencies)
and the presence of additional light-waves are shown respectively in \cite{Noi1,Noi2} and \cite{Noi3}.
The former implies that the electromagnetic energy flux and the phase velocity of an electromagnetic 
wave through the medium propagate in opposite directions \cite{SpatNeg, Metamat}; the latter means
that, even if isotropic, a medium supports multiple electromagnetic waves with the same frequency but different wave-vectors \cite{ALW}.
Following these results, various authors investigated these topics in a wide variety of different setups, see for example \cite{OtherNegRef}. 
In particular \cite{DJD} provides an analysis of the actual signature of the abovementioned phenomena in condensed matter systems and it proposes suitable 
experiments to measure such effects.

Even though much of the previously stated theoretical progress about NR and ALW was obtained within the framework of strongly coupled plasmas,
it turned out that the qualitative behavior of the discovered phenomena has a more general valence and 
it constitutes a universal property of electrically charged hydrodynamical systems \cite{Noi2}. 
The main purpose of the present paper is to go beyond the regime described by hydrodynamics 
in systems characterized by strong coupling physics. 
In particular, we will show that the coupling of the external electromagnetic field with 
the tower of quasi-normal modes of strongly coupled plasmas leads to the presence of various 
electromagnetic modes which present different properties. For instance, it is important to stress 
that we found negative refraction with low dissipation for a transverse non-hydrodynamical mode.

Gaining insight on the strongly coupled dynamics is 
crucial in order to describe some very interesting phases of matter which have recently been realized both 
in high energy physics experiments (e.g. the quark gluon plasma (QGP), see for example \cite{CasalderreySolana:2011us}) 
and in condensed matter systems (e.g. high-$T_c$ superconductors, see for example \cite{Zaanen:2010yk} for a recent review%
\footnote{To have a recent and brief introduction on the gauge/gravity approach to describe strongly coupled superconductors
see for instance \cite{Musso:2014efa}.}). 
Moreover, the existence of strong interactions among the various resonances in some of the artificially engineered meta-materials, 
the so called stereo-meta-materials \cite{Stereo}, provides yet another interesting 
phenomenological stimulus to investigate the electromagnetic behavior and response 
of strongly coupled systems. 

The idea of performing an analysis beyond the hydrodynamical regime aims at studying
the electromagnetic wave modes that are not controlled by the hydrodynamical universal behavior
accounted by the equations of fluid dynamics%
\footnote{A correlated question which is very interesting and treatable within the framework of
the gauge/gravity correspondence is the relation among thermalization and hydrodynamization. 
To have a recent account on this topic we indicate \cite{Janik:2014kfa,Stricker:2014cma} and references therein.}. 
Particular attention is then paid to observe whether 
the exotic phenomena arising already in hydrodynamics are present also beyond the 
long wavelength and long time approximations and what are their properties in this regime. 

It is well known that the dynamics at strong coupling is usually very difficult to study and,
outside the regime constrained by hydrodynamics, it is quite difficult to provide universal predictions.
However, some interesting theoretical progress has been recently obtained thanks to techniques coming from string theory. 
These new methods exploit a conjectured duality between a strongly coupled quantum field theory and a weakly 
coupled gravity theory; such duality is usually referred to as gauge/gravity correspondence \cite{AdSCFT}. 
The gauge/gravity duality (and similar correspondences) allows us to quantitatively study the correlation functions of specific models
and, at the same time, to understand better many interesting and generic qualitative behaviors associated to various (universality) classes 
of strongly coupled systems. Insights from the gauge/gravity correspondence 
appeared to be relevant both to high energy physics and condensed matter (see for instance 
\cite{KovOrov} to have two paradigmatic examples) especially in relation to the characterization of the phase structure 
and transport properties.
Within the gauge/gravity correspondence framework, quantitative computations of expectation values and correlation functions of
strongly coupled quantum systems at finite temperature are reduced to classical gravity calculations in specific 
black hole backgrounds (which are actually finite temperature gravitational systems). 
The thermodynamical observables of the strongly coupled system in equilibrium are mapped to 
classical properties of the dual gravity model. A semi-classical study of the equations of motion 
for the fields of the dual gravity model unveils both the equilibrium thermodynamics and the linear response
(i.e. the slightly out of equilibrium physics). 
The latter being obtained considering linearized equations of motion
(equipped with appropriate boundary conditions) describing small fluctuations around the gravitational black hole background.

To our aim, the benefit of the gauge/gravity framework is twofold: first of all it allows us to study correlation functions
for any value of the frequency and wave-vector in a well defined and complete setup (essential to go beyond hydrodynamics); 
secondly, it provides interesting hints on generic responses of strongly coupled systems beyond the hydrodynamical 
regime.

A possible phenomenological application of our results is to QGP physics.
The neutral plasma that we study could indeed mimic the QGP plasma in a regime dominated by the temperature where the finite chemical potential is neglected.
In particular, we stress that the electrodynamical, non-hydrodynamical modes described in this paper 
have smaller wavelengths when compared to the hydrodynamical modes studied in \cite{Noi2}; hence they probe smaller
distances which could be relevant for actual QGP samples produced in experiments.

The paper is organized as follows. In Section \ref{scene} we briefly review 
the linear response theory needed to study the electromagnetic properties of a medium.
Section \ref{results} contains our main results obtained by studying
 the first transverse electromagnetic waves through the strongly coupled plasma.
We show in particular the generic presence of ALW and we demonstrate the existence of a 
propagating electromagnetic mode with negative refraction and very small dissipation.
In Section \ref{mod} we describe the actual calculations venturing
into technical details of both the gauge/gravity setup and the peculiarities 
of the necessary renormalization procedure. Section \ref{nume} is dedicated to 
the description of our numerical and semi-analytical methods and the checks that our 
results have undergone. In Section \ref{QGP} we discuss very briefly some phenomenological 
applications of our study to the QGP. In Section \ref{long} we show the results of an analogous analysis
regarding the longitudinal light-wave modes. Eventually, in Section \ref{con} we conclude
stressing the significance of the results and possible future prospects. The appendices 
contain a characterization of the  
quasi-normal modes of the plasma (both transverse and 
longitudinal) and the details about the computation regarding the longitudinal light-wave sector.

\section{Setting the scene}
\label{scene}

The electromagnetic response of a system in local equilibrium to an external electromagnetic perturbation is described by the 2-point
retarded correlation function of the electromagnetic current \cite{Dress}.
If, as in the case we are considering, the system has spatial dispersion, the response function depends on the space positions where the currents are evaluated.
In the particular case of a homogeneous and isotropic system, the correlation functions depend only 
on the distances and they can be further decomposed into a transverse and a longitudinal part with 
respect to the wave-vector $q$ of the external perturbation. 
In this case, as shown for example in \cite{SpatNeg,Landau}, it is possible to describe the macroscopic electromagnetic
properties of the system with three fields: $D$, $E$ and $B$. Moreover, the linear response of the medium 
(in Fourier space) is accounted for by a single tensorial function depending 
both on the frequency $\omega$ and the wave-vector $q$, the so-called permittivity tensor $\epsilon_{ij}( \omega, q)$, 
\begin{equation}\label{DEeps}
D_i =\epsilon_{ij}(  \omega, q)E_j \ .
\end{equation}
As a consequence of the isotropic assumption, the permittivity tensor is expressed in terms of only two scalar functions 
\begin{equation}\label{DEeps}
 \epsilon_{ij}(  \omega, q) =\epsilon_{T}(  \omega, q)  \left(  \delta_{ij} - \frac{q_i q_j}{q^2} \right) +  \epsilon_{L}(  \omega, q) \frac{q_i q_j}{q^2}\ ,
\end{equation}
describing the transverse and longitudinal response with respect to the spatial momentum $q$ of the 
external perturbation.

The 2-point retarded correlation functions, denoted as ${\cal G}_T(\omega,q)$ and ${\cal G}_L(\omega,q)$ for the transverse and the longitudinal part respectively, 
characterize completely the permittivity tensor:  
\begin{equation}\label{eps_tra}
 \epsilon_{T,L}(\omega,q) = 1 - 4 \pi e^2 \,  \frac{ {\cal G}_{T,L}(\omega,q)}{\omega^2}\ ,
\end{equation}
where $e$ is the electromagnetic coupling constant.
The knowledge of ${\cal G}_{T,L}(\omega,q)$ allows us to study the propagation and dissipation
of the electromagnetic waves inside the medium at the linear level. Indeed, the Maxwell equations 
for the electromagnetic field provide the dispersion relation between $q$ and $\omega$ both for the transverse and the longitudinal channel%
\footnote{In a spatially dispersive medium also the longitudinal electromagnetic waves do propagate.}. 
The Maxwell equations for the transverse and longitudinal
part of the electric field $E$ imply the dispersion equations:
\begin{equation}\label{eps_traK}
 \epsilon_T(\omega,q)\ = \frac{q^2}{\omega^2}\ ,  \qquad \epsilon_L(\omega,q)\ =0 \ ,
\end{equation}
which admit solutions $q(\omega)$ corresponding to possible electromagnetic waves through the medium%
\footnote{A description of the electromagnetic waves in terms of functions $\omega(q)$ is possible as well; 
see for instance \cite{Ginzburg}}. The information about the propagation and dissipation of an electromagnetic
mode through the system is encoded in the refractive index which is defined as $n^2(\omega)=q^2(\omega)/\omega^2$
considered on the dispersion relation of the mode under study. 

Albeit we study both the transverse and the longitudinal electromagnetic sectors, we focus mainly
on the detailed analysis of the transverse sector (whose properties are more interesting to us).
An analogue detailed treatment of the longitudinal sector is reported in Appendix \ref{lon_det}.

\section{Results}
\label{results}

In this Section we describe the main results of our analysis. We focus on the refractive indexes of the first 
electromagnetic transverse modes inside a strongly coupled plasma composed of electrically charged constituents but globally neutral. 
The plasma we study is the finite temperature state of a relativistic and conformally invariant particle 
physics theory composed of fermions and bosons\footnote{Further details of the actual theory and setup will be 
provided in Section \ref{mod}.}.
The choice of this particular kind of plasma is due to the existence of very powerful theoretical tools 
(provided by the gauge/gravity correspondence) to study its strongly coupled dynamics. 
Indeed the present analysis can be read as a first example to investigate the potentiality of the gauge/gravity 
correspondence to explore exotic electromagnetic phenomena of strongly coupled, finite temperature systems
beyond the hydrodynamic regime.

The plasma is assumed to be homogeneous and isotropic\footnote{These are assumptions that make our analysis easier. 
Non-homogeneous or non-isotropic cases, charged plasmas, superconducting phases and non-relativistic theories are
very interesting to investigate too and are left for future work.}.
Previous studies have shown that plasmas described with the gauge/gravity correspondence have in general an 
infinite set of discrete quasi-normal (complex valued) 
self-frequencies \cite{Quasi,Amado:2007yr,Amado:2008ji}, that appear
as poles in the two-point retarded correlation function of the currents.
In this paper we are interested to couple these quasi-normal modes of the plasma to an external electromagnetic 
field and study the possible electromagnetic modes supported by the system. It is important to observe that, 
even if the quasi-normal modes of the medium appear as poles of the correlation functions, due to high dissipation,
the quasi-particle approximation is usually not valid. 
This implies that at any frequency the response function can not be approximated by a single resonance.
Indeed, the self-frequencies of the system are in general complex and the amount of dissipation and propagation 
of the quasi-normal modes is usually comparable%
\footnote{Note that only in this paragraph we refer to the framework in which $\omega$ is complex and $q$ is real.}.

\begin{figure}[t]
\centering
\includegraphics[width=78mm]{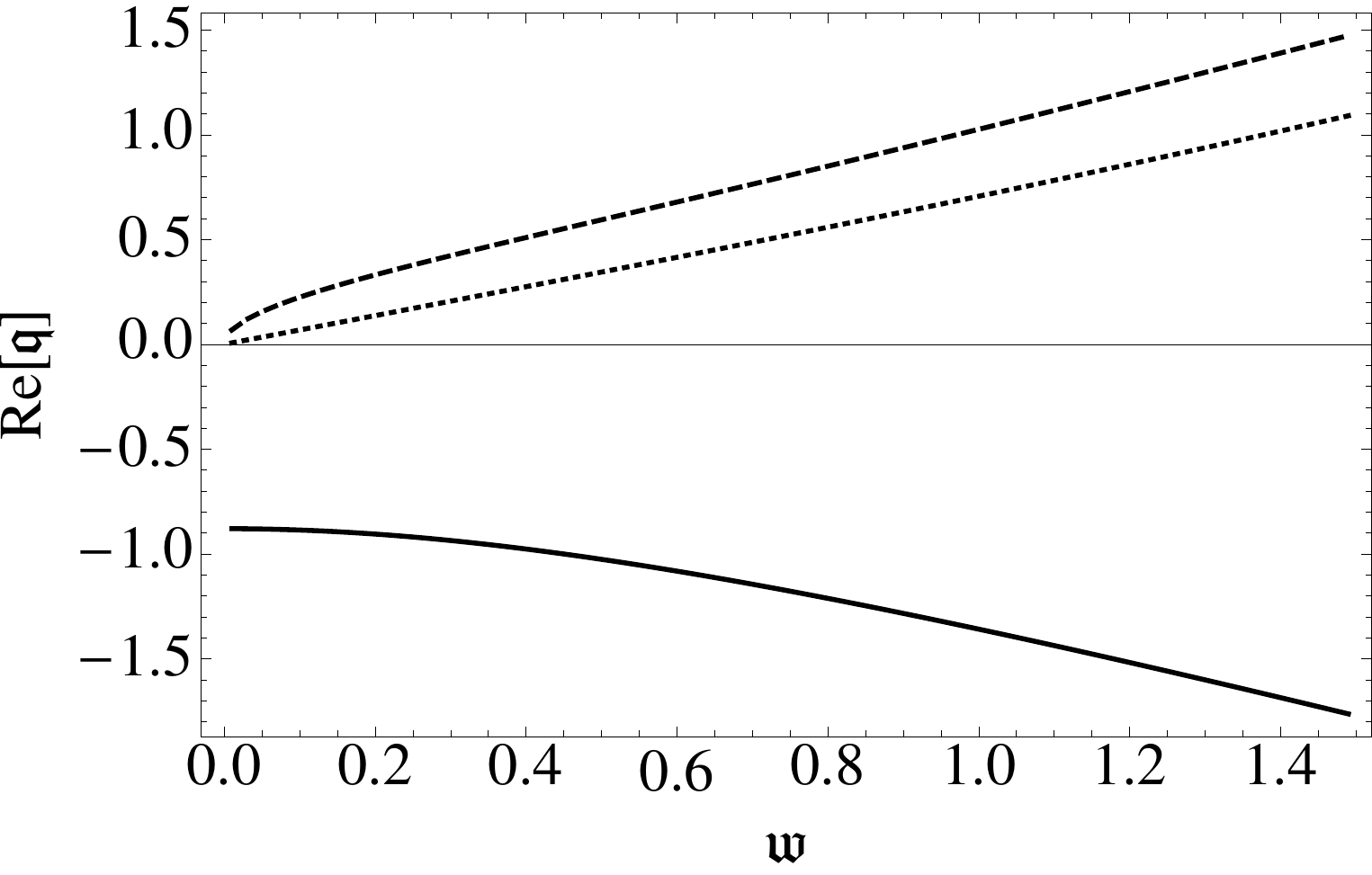} 
\includegraphics[width=78mm]{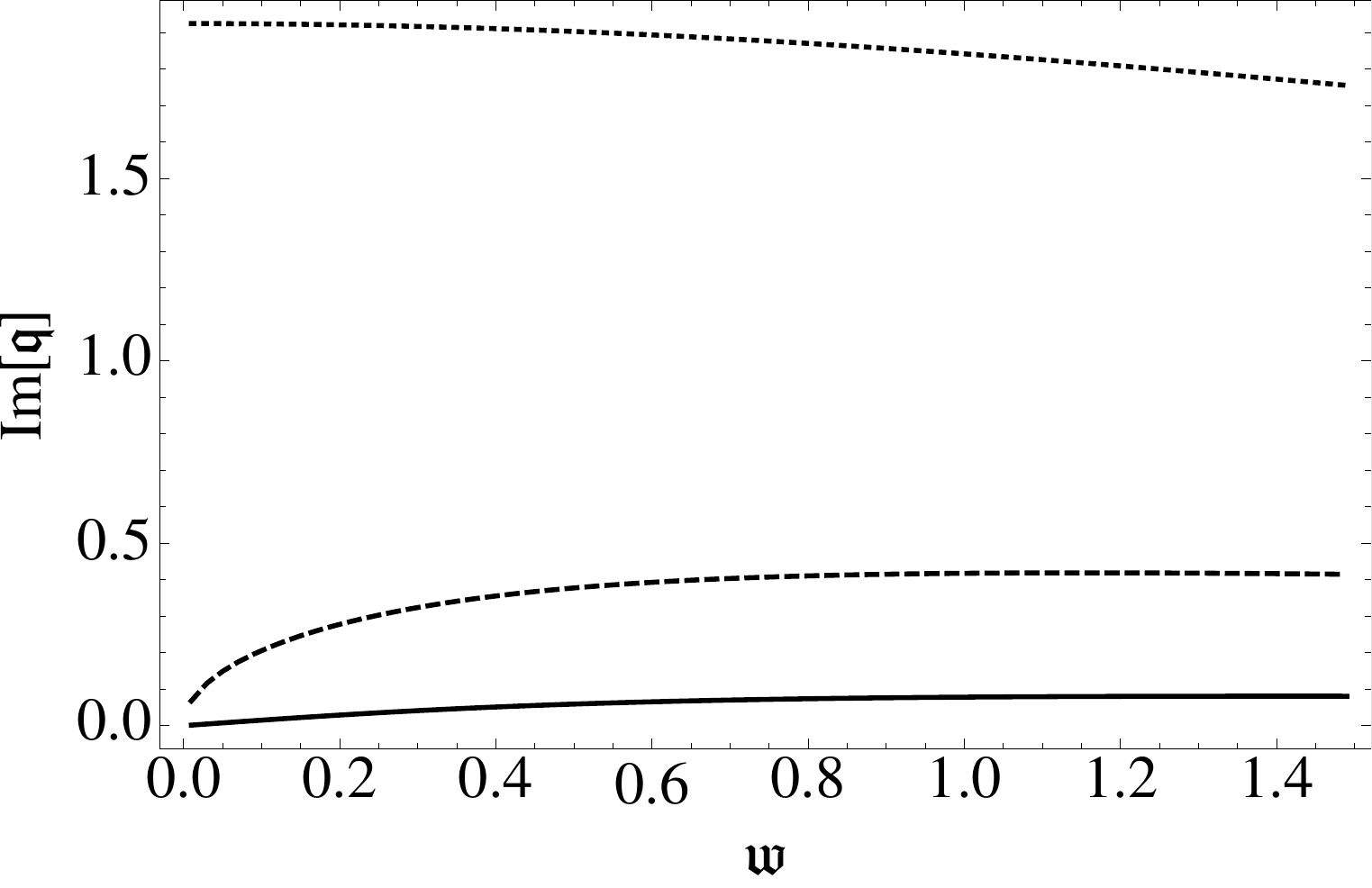} 
\caption{Real (left) and imaginary (right) parts of the rescaled wave-vector ${\mathfrak q}$ 
as a function of the rescaled frequency ${\mathfrak w}$ for the first three transverse electromagnetic modes of the 
neutral plasma under study.}
\label{q}
\end{figure}

\begin{figure}[t]
\centering
\includegraphics[width=78mm]{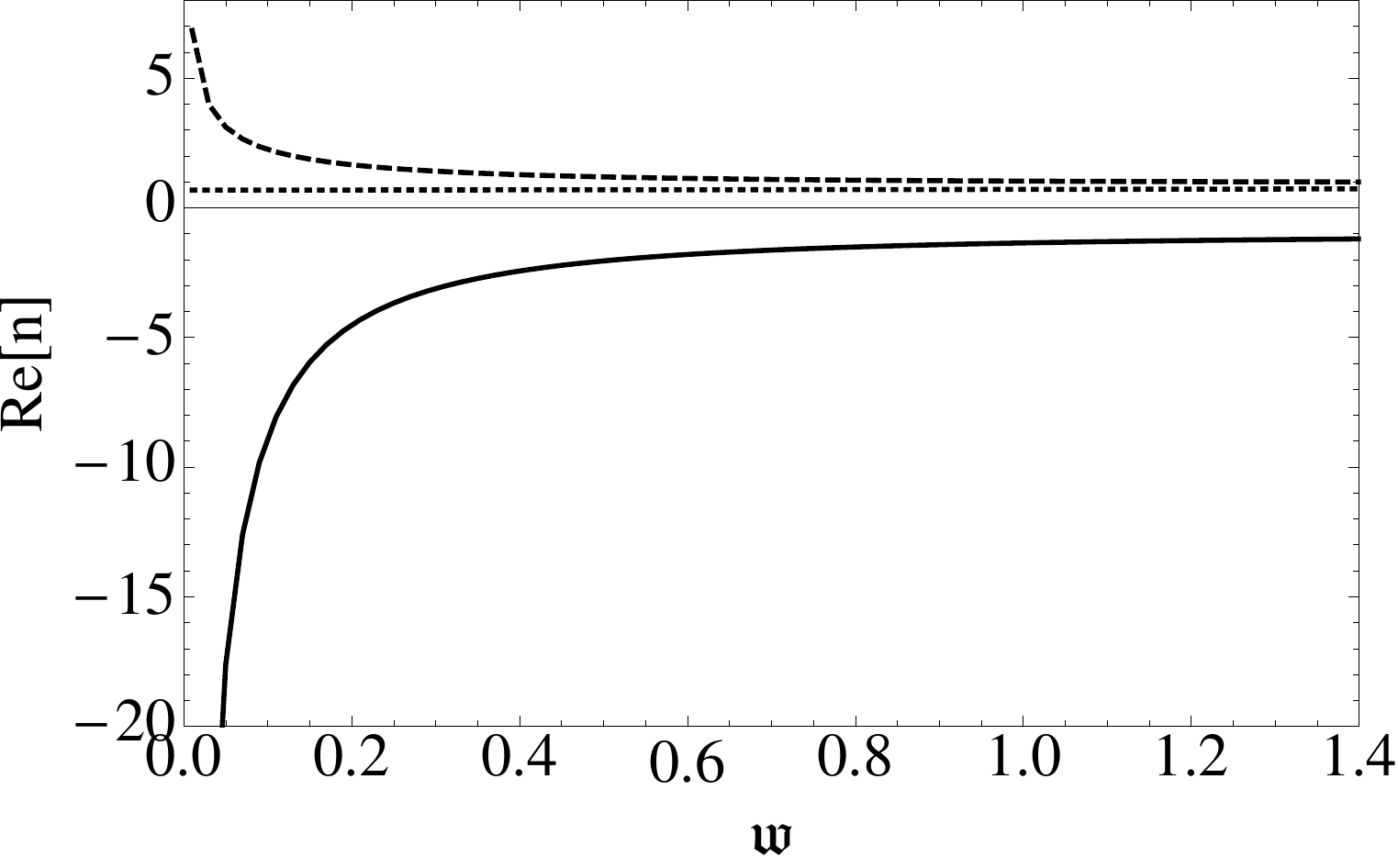} 
\includegraphics[width=78mm]{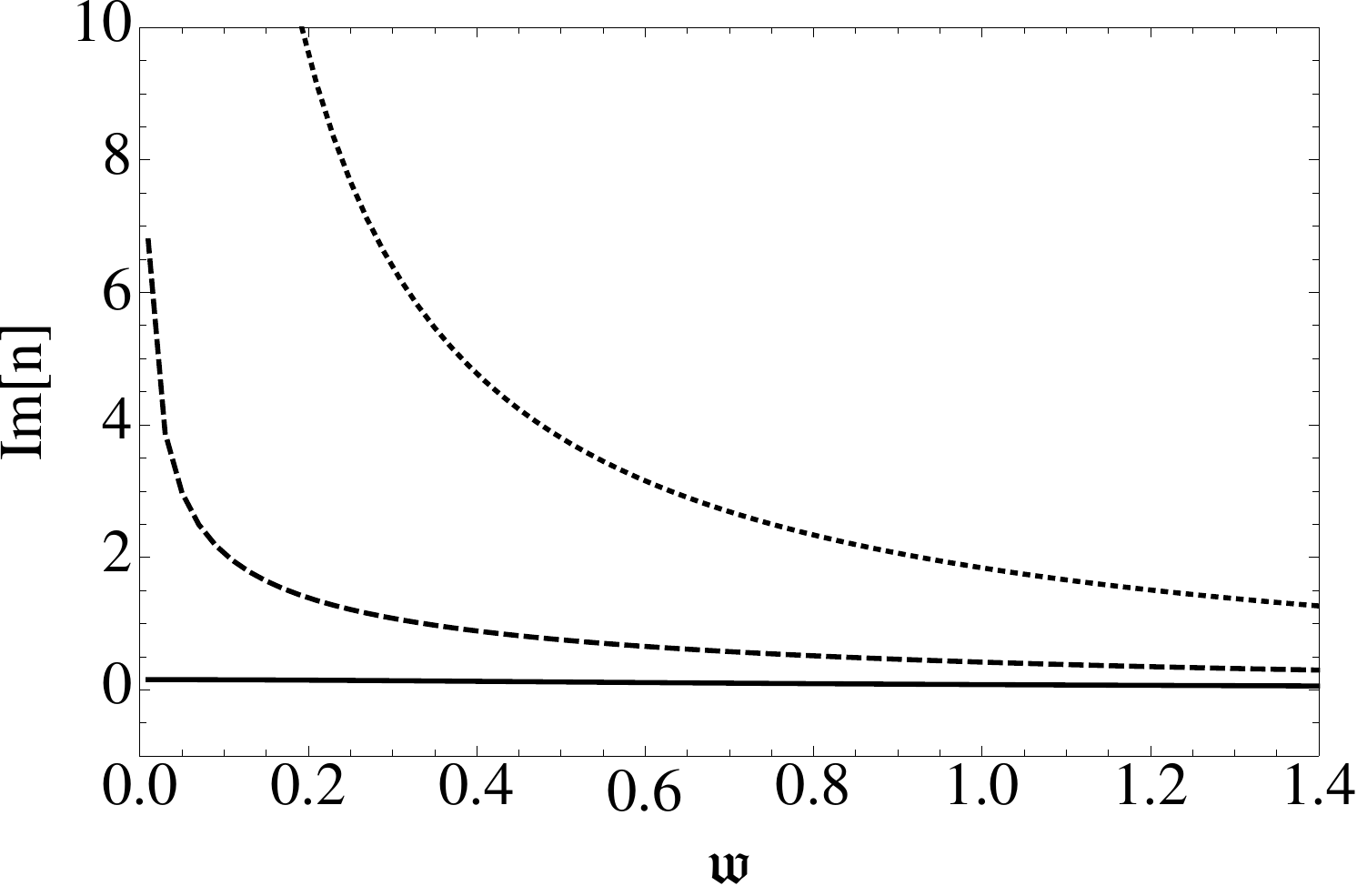} 
\caption{Real (left) and imaginary (right) parts of the refraction index $n$
as a function of the rescaled frequency ${\mathfrak w}$ for the first three transverse electromagnetic modes of the 
neutral plasma under study.}
\label{n}
\end{figure}

\begin{figure}[t]
\centering
\includegraphics[width=78mm]{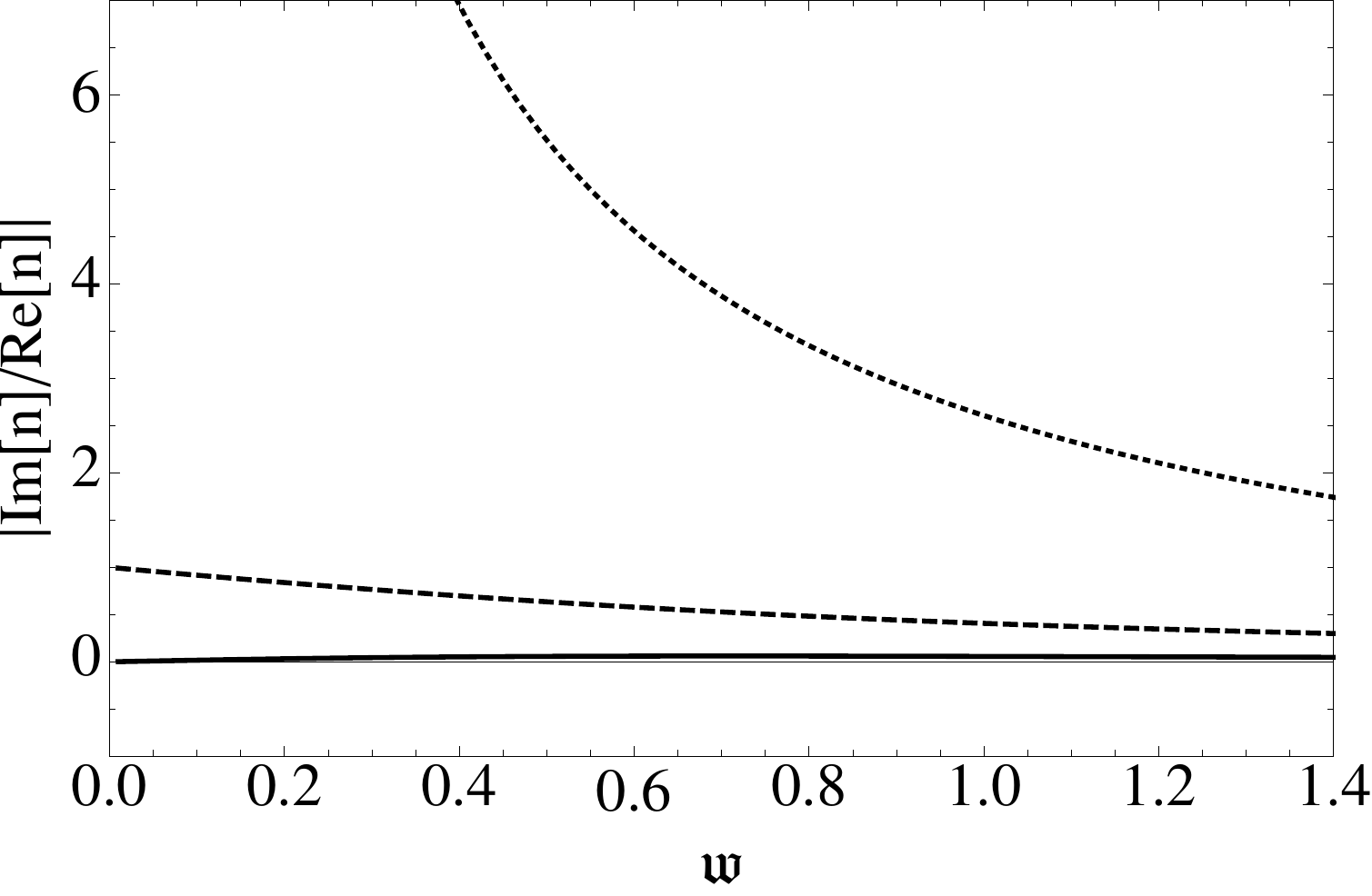} 
\includegraphics[width=78mm]{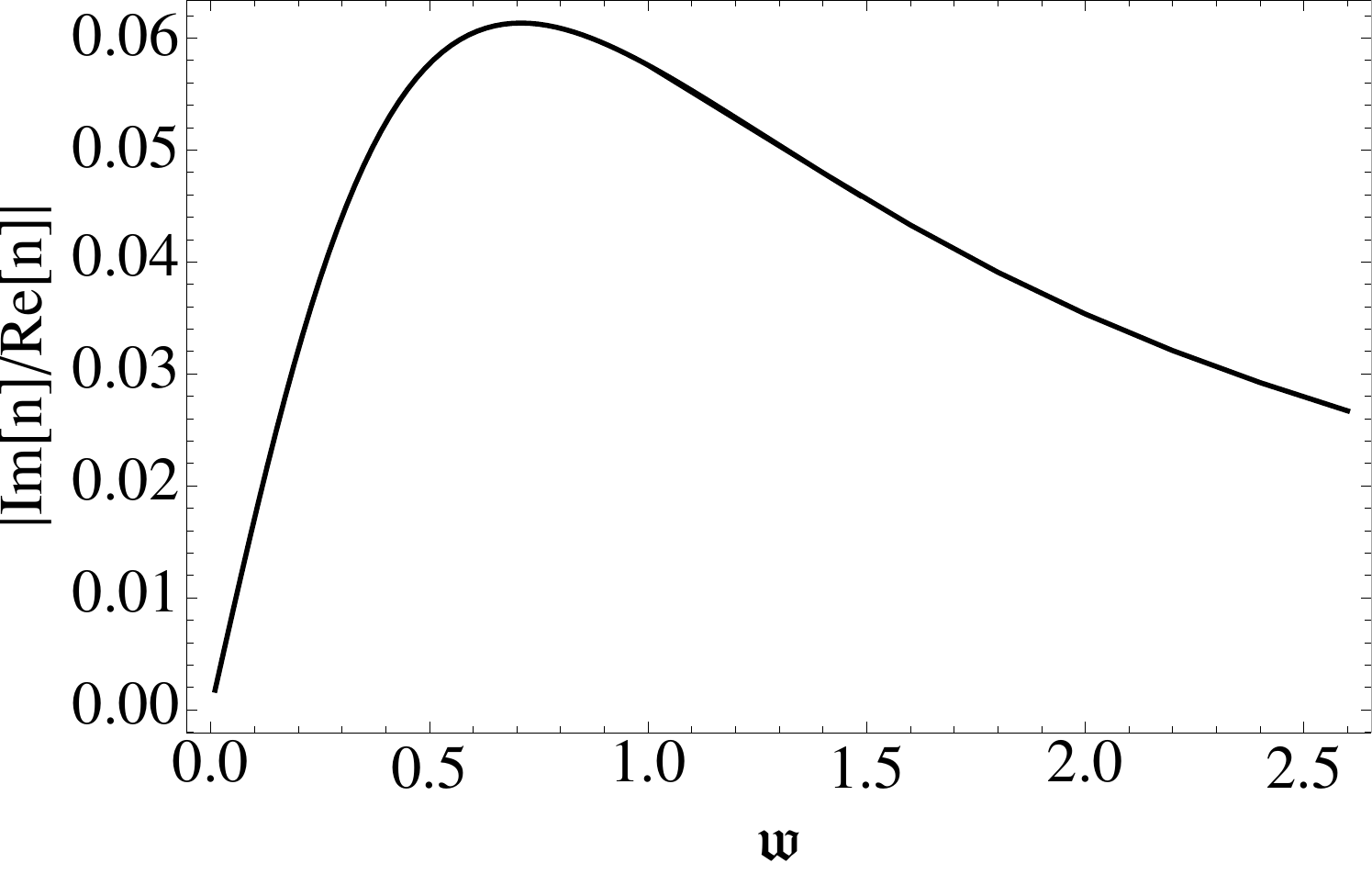} 
\caption{Absolute value of the ratio of the dissipation/propagation ratio
for the first three electromagnetic modes of the plasma (left). Rescaled plot 
of the negatively refractive mode (right).}
\label{prop}
\end{figure}

The propagation and dissipation of an electromagnetic wave is described by the corresponding refractive index: 
the real part of the index accounts for the propagation of the wave while the imaginary part describes 
its attenuation. 
We focus here on the transverse modes of the electromagnetic field. The analysis of the longitudinal channel
is contained in Section \ref{long} and Appendix \ref{lon_det}. 
A study of the first three transverse modes (i.e. those associated to the lowest wave-numbers) reveals the rich structure of the 
electromagnetic response of the medium.
We have analyzed the real and imaginary parts of their refractive index and of the wave-vector $q$ as functions of the frequency, and, in order to characterize the 
propagation/dissipation rate of the aforementioned modes, we have studied the absolute value of the ratio $\text{Im}(n)/\text{Re}(n)$%
\footnote{Since $n=q/\omega$, studying the ratio $\text{Im}(n)/\text{Re}(n)$ is equivalent to studying 
$\text{Im}(q)/\text{Re}(q)$; recall indeed that $\omega$ is here a real quantity. Therefore,
the ratio $\text{Im}(n)/\text{Re}(n)$ can be read as the comparison between the wavelength and the
characteristic attenuation length of the signal.}.

The results are presented in Figures \ref{q}, \ref{n} and \ref{prop}.
For reasons that will become clear in the next Sections we  
plot the results using the rescaled wave-vector ${\mathfrak q}$ and the rescaled frequency ${\mathfrak w}$, 
obtained by dividing the wave-vector number and the frequency by the temperature of the system 
\begin{align}\label{Twq}
\mathfrak{w} = \frac{\omega}{2 \pi T}\ , 
\ \ \ \ 
\mathfrak{q} = \frac{q}{2 \pi T}\ .
\end{align}
A quite interesting picture arises from our analysis of the electrodynamic response of the plasma. The system 
can indeed  support various electromagnetic modes, some of them evanescent and other propagating, with different peculiar characteristics.
The wave-vector of one of the modes vanishes for vanishing frequency while those corresponding
to the other two modes reach a finite value at zero frequency (one purely real and the other purely
imaginary). The presence of a finite value of $\mathfrak{q}$ at null frequency (regardless of its real, imaginary or complex character) 
is due to the coupling of the electromagnetic waves to non-hydrodynamical quasi-normal modes of the plasma.
The mode presenting an imaginary $\mathfrak{q}$ at zero frequency is very dissipative in the low-frequency region.
On the contrary, the one presenting a real $\mathfrak{q}$ at zero frequency propagates with very small damping.
It is very interesting to observe that the real part of the refractive index of this last propagating electromagnetic mode is negative 
implying that it is a negatively refractive mode, i.e. its energy flux and phase velocity are directed in opposite directions%
\footnote{The system we are studying is a passive medium which can only dissipate and not source energy. This implies that the 
sign of the imaginary part of the refractive index agrees with the direction of propagation of the electromagnetic 
energy flux in the medium \cite{mccall}. Intuitively, in a passive medium, a wave propagates in the direction in which it is damped.}.

In some previous studies \cite{Noi1,Noi2} it has been proven that for charged plasmas in the hydrodynamic regime
it always exists a mode with negative refraction for small enough frequency, while there is no negative refraction for
globally uncharged plasmas.
In the present paper we extended the investigation beyond the hydrodynamical modes and we proved that
globally neutral, strongly coupled plasmas can support modes with negative refraction too. 
Moreover, in \cite{Noi3}, the presence of multiple electromagnetic modes 
was shown to be generic for the hydrodynamical regime. Here we extended such analysis to entail also 
non-hydrodynamical modes and we showed that ALW still appear to be present. In this extended context it is natural to relate the presence
of ALW to the coupling of the electromagnetic radiation to the infinite tower of quasi-normal modes.

Note that we studied the modes at frequencies $\mathfrak{w} \lesssim 1$. 
In this regime a neat hierarchy between the modes
arises from their dissipation over propagation ratio. From Figure \ref{prop} one can observe that
the negative refraction mode dominates the signal propagation. In line with the collected numerical data
one expects that, increasing the frequency, more electromagnetic waves propagate significantly. To have 
a faithful picture of the plasma response (in terms of propagating waves) at higher frequency it is 
therefore necessary to include higher modes into the analysis%
\footnote{Our methods allow also a study at higher frequency though a heavier numerical 
effort is required. For instance, the negatively refracting mode has been studied up to $\mathfrak{w}\sim 2.5$,
see the right plot of Figure \ref{prop}.}. The negative refraction phenomenon of the plasma at hand is then likely to 
be significant in the low-frequency regime. Of course a careful study of the boundary conditions would be required to 
actually understand which modes are excited 
in a specific experimental setup.
In this paper we concentrate on the bulk properties of the medium and we leave this problem for future investigation.

As a final comment, let us mention that in this Section we presented the results without making reference to the specific value
of the temperature or other physical parameters of the system. Let us anticipate that the qualitative behavior 
of the system is robust and valid in a wide range of parameters. A precise treatment of the quantitative predictions of 
our model requires a careful analysis of the regularization/renormalization procedure which we postpone to later sections.

\section{Stringy setup and gauge/gravity computation}
\label{mod}

In this Section we provide some details about the actual class of systems we consider, 
the computations we performed and the associated dual gravitational setup. 
The material collected in the first two Subsections is quite standard and describes how to perform computations in the 
gauge/gravity correspondence at finite temperature. Nevertheless, we decided to report it in some detail 
since some subtleties of the computation are relevant for our analysis concerning the electromagnetic modes.
Experts in gauge/gravity correspondence could however speed through the discussion below and go to Subsection \ref{lw}.

As we have previously stated we consider the electromagnetic response of strongly coupled neutral plasmas. 
The gauge/gravity correspondence allows us to describe the quantum dynamics at strong coupling of 
a gauge field theory defined on 4-dimensional Minkowski space-time in terms of a classical gravitational model living in a 
5-dimensional Anti-de Sitter space-time ($AdS_5$).
Furthermore, the gauge/gravity correspondence maps the finite temperature phase of a quantum field theory into 
a finite temperature gravity configuration: namely a black hole solution which becomes asymptotically $AdS_5$ at large distances ($AdS$ black hole).

We do not restrict ourselves to any particular model, but we study instead the 
generic behavior of the universal sector of a strongly coupled theory containing the energy momentum 
tensor $T_{ \mu  \nu}$ and the $U(1)$ conserved current $J_{ \mu}$. 

In the context of the gauge/gravity correspondence, these two minimal ingredients are mapped respectively to a 5-dimensional metric $g_{mn}$ and a 5-dimensional 
vector field $A_{m}$, whose dynamics is encoded in the 5-dimensional Maxwell-Einstein action\footnote{Our notations are coherent with \cite{Hartnoll:2009sz,Argurio:2013uba}.
}
\begin{equation}\label{action}
 S = - \frac{N^2}{32 \pi^2} \int d^5x \sqrt{-g} \left( R - 2\Lambda \right) 
 + \frac{N^2}{16 \pi^2}  \int    d^5x\; \sqrt{-g}\;\frac{1}{4} F_{mn} F^{ mn} \ ,
\end{equation}
where $\Lambda$ is the cosmological constant and $N$ is, in the gravity model perspective, a normalization constant. 
The gravitational constant has been related to the $AdS_5$ radius of curvature $L$ and the string constant $\alpha'$
which have been both put to 1.

The ground state at finite temperature $T$ and zero charge density of a strongly coupled quantum field theory is dually
described by the neutral $AdS_5$ black hole solution of the action \eqref{action} specified by the following bulk metric written in the Poincar\'e patch
\begin{equation}\label{bg}
 ds^2 = \frac{\pi^2 T^2}{u} \left[-f(u) dt^2 + dx_i dx^i\right] + \frac{1}{4 u^2 f(u)} du^2\ ,
\end{equation}
where $f(u) = 1-u^2$.  The horizon of the black hole 
is located at $u=1$, while the $AdS_5$ conformal boundary corresponds to $u=0$.

In the gauge/gravity framework, the study of the fluctuations of the 5-dimensional photon $A_{m}$ 
on the background specified by \eqref{bg} provides information on the retarded 2-point correlation 
functions of the $U(1)$ current $J_{ \mu}$ of the dual $(3+1)$-dimensional quantum field theory \cite{Son:2002sd}. 
As already stated, the knowledge of the retarded 2-point correlation function of the currents 
is the required element to discuss the electromagnetic properties of such strongly coupled plasmas.

\subsection{Vector transverse fluctuations}

We focus on the fluctuations of the 5-dimensional bulk photon $A_m$.
We fix partially the 5-dimensional gauge invariance imposing $A_u = 0$;
note that this $5$-dimensional ``radial'' gauge condition still leaves the usual $4$-dimensional gauge freedom 
for the remaining components of the vector potential $A_\mu$.
In other words, we study the fluctuations of the bulk gauge field along the $4$-dimensional
space-time directions parametrized with $\mu=t,x,y,z$ where the dual theory is 
actually defined. Moreover, we consider the fluctuations
which are transverse to the direction of propagation, namely the fluctuations that are
perpendicular to the spatial wave-vector $\vec{q}$. 

We decompose $A_\mu$ in $4$-dimensional Fourier modes: 
\begin{equation}\label{FAi}
 A_\mu(u,t,\vec{x}) = \int \frac{d\omega d^3q}{(2\pi)^4}   e^{  -i (\omega t - \vec{ q }  \cdot \vec{x} )} A_\mu (u,\omega,\vec{q}) 
 \ .
\end{equation}
Since the system is invariant with respect to spatial rotations, we can in general take the spatial momentum 
of the fluctuations to be along the $z$ direction so that the $4$-momentum is
$k^\mu = (\omega,0,0,q)$. 
The transverse fluctuations are therefore those along the $x$ and $y$ directions
and we introduce the label $\alpha = x,y$ to refer to this transverse space.
Note that, as we have cylindrical symmetry around the $z$ direction, we can focus on the gauge field fluctuations 
along $x$ without spoiling the generality of the treatment.
Specifically, we adopt the following ansatz for the Fourier components
\begin{equation}\label{ans}
 A_\alpha(u,\omega,q)\, dx^\alpha = \phi(u,\omega,q)\, dx\ .
\end{equation}
Plugging this ansatz into the Maxwell equation derived from the action \eqref{action} and considered on the background (\ref{bg}),
we obtain 
\begin{equation}\label{transverse}
 \phi'' + \frac{f'(u)}{f(u)} \phi' + \frac{\mathfrak{w}^2-\mathfrak{q}^2 f(u)}{u f(u)^2} \phi = 0 \ ,
\end{equation}
which, by cylindrical symmetry, is valid for the fluctuations along a generic transverse direction.
The $u$, $q$ and $\omega$ dependence of $\phi$ is understood,
the prime indicates the derivative with respect to the radial coordinate $u$; we also adopted the dimensionless frequency $\mathfrak{w} = \omega/(2\pi T)$
and momentum $\mathfrak{q}=q/(2\pi T)$ already introduced in \eqref{Twq},
where $T$ is the temperature of the dual field theory%
\footnote{Equations (\ref{Twq}) provide a meaningful definition 
of the concept of small or large frequency and wave-vector. Recall that the microscopic theory that we describe is conformally invariant at 
zero temperature; hence the temperature is the only scale against which we can 
meaningfully compare the magnitude of the frequency and wave-vector.} or, equivalently, of the $AdS$ black hole in (\ref{bg}). 

A near-horizon analysis of Equation (\ref{transverse}) yields the following asymptotic behavior for the 
field $\phi$
\begin{equation}\label{IR_exp}
 \phi = (1-u)^{\pm i  \mathfrak{w}/2} \left[ a_{\pm} + b_{\pm} (1-u) + c_{\pm} (1-u)^2 + ...\right]\ .
\end{equation}
The generic solution to the differential equation is given by the superposition of an in-going and an out-going solutions. 
We select the in-going solution imposing the condition $a_{+}=0$ in accordance with the idea that nothing can be emitted by the black hole horizon at the classical level.
This choice corresponds to compute retarded correlators in the dual field theory.
As we are confronted with a second order differential equation we need to impose one further boundary 
condition. Relying on the linearity of Equation \eqref{transverse} we choose $a_{-}=1$; as we will shortly explain,
we are eventually interested in ratios like $\phi'/\phi$ which are completely insensitive to this specific choice.
As we will describe in detail in Section \ref{nume}, once equipped with these horizon boundary conditions, 
we are able to solve the differential problem propagating the solution from the horizon.

We now turn the attention to the near-boundary region corresponding to small $u$. 
This asymptotic UV analysis is necessary to extract from our model the physical quantities we are interested in.
A near-boundary term-wise study of the equation of motion \eqref{transverse} near the $AdS$ boundary at $u=0$
shows that the asymptotic expansion of the field $\phi$ is there
\begin{equation}\label{UV_phi}
 \phi = \phi_0 + u\, \phi_1 + u \ln(u)\, \tilde{\phi}_1
 + u^2 \phi_2 + u^2\ln(u)\, \tilde{\phi}_2 + ...\; ,
\end{equation}
where the $\phi_i$ and $\tilde{\phi}_i$ coefficients are independent of the $u$ coordinate, and satisfy%
\footnote{The background specified in \eqref{bg} is asymptotically $AdS_5$ and 
a similar study of the near-boundary behavior of the transverse vector fluctuations
on pure $AdS_5$ leads to the same relations \eqref{rel} among the first UV coefficients.}
\begin{eqnarray}\label{rel}
 (\mathfrak{w}^2 - \mathfrak{q}^2) \phi_0 + \tilde{\phi}_1 &=& 0\ , \\
 2 \tilde{\phi}_2 + (\mathfrak{w}^2 - \mathfrak{q}^2) \tilde{\phi}_1 &=& 0\ , \\
 2 \phi_2 + 3 \tilde{\phi}_2 + (\mathfrak{w}^2 - \mathfrak{q}^2) \phi_1 &=& 0\ .
\end{eqnarray}
Note that, since the equation of motion is a second order differential equation,
we can express all the coefficients of this term-wise solution as functions of the first two
coefficients $\phi_0$ and $\phi_1$.

\subsection{Correlation functions and holographic renormalization}
\label{Green}

In this Section we follow the gauge/gravity prescription to compute the retarded two-point 
correlator of the transverse currents of the strongly coupled plasma.
Indeed, once we found the solution to Equation (\ref{transverse}) with the boundary condition explained in the previous section, 
we can obtain the retarded correlator of the current \cite{Son:2002sd}. More specifically, as
it is usual in quantum field theory, one can derive the correlators through functional differentiation with respect
to the sources; from the gravity model standpoint, the procedure corresponds to functionally 
differentiate the on-shell bulk action with respect to the boundary value of the 
fluctuating fields.

The on-shell gauge field action for the transverse vector field 
(described by the solution $\phi$ of the previous Section) becomes
\begin{equation}\label{os_lag_tra}
 \frac{(N T)^2}{16} \int \frac{d\omega d^3q}{(2\pi)^4}\ f(u) (\phi')^2\ ,
\end{equation}
which, after integration by parts and having considered the equation of motion, leads to the boundary term
\begin{equation}\label{bou_ter}
 \frac{(N T)^2}{16} \left[ f(u) \phi \phi' \right]' \ .
\end{equation}
This latter is to be integrated on the boundary manifold. Interestingly, the on-shell 
action reduces to boundary terms and only the contribution from the $AdS$ conformal 
boundary at $u=0$ is not vanishing. The contribution from the horizon at $u=1$ actually vanishes because 
$f(u)$ is there null. As a consequence, we just focus on the contribution to the primitive of \eqref{bou_ter}
at asymptotically small $u$; we expand the field $\phi$ and its derivative as in \eqref{UV_phi} discarding the terms that vanish at $u=0$
and we obtain
\begin{equation}\label{asy_bou}
 \frac{(N T)^2}{16}\ \int \frac{d\omega d^3 q}{(2\pi)^4}\ \phi_0 \left[ \phi_1 - {\phi}_0 (\mathfrak{w}^2 - \mathfrak{q}^2) (1 +\ln(u)) \right]\ .
\end{equation}
It is important to observe that there exists a logarithmically diverging term. 
This signals the need to renormalize the model. 
The occurrence of a large-volume divergence in the gravity theory living on the asymptotically $AdS$ geometry corresponds 
(through gauge/gravity correspondence) to the UV divergences of the dual quantum field theory. 

Let us describe the renormalization procedure \cite{RenAdSCFT} of the model at hand.
At first we regularize the on-shell action considering a small $u=\epsilon$ cutoff,
\begin{equation}
 S_{\text{reg}} =\frac{(N T)^2}{16} \int_{u=\epsilon} \frac{d\omega d^3 q}{(2\pi)^4}\ \phi_0 \left[ \phi_1 - \phi_0 (\mathfrak{w}^2 - \mathfrak{q}^2) (1 + \ln(\epsilon))\right]\ .
\end{equation}
We then add an appropriate boundary counter-term
\begin{equation}\label{ct}
 S_{\text{c.t.}} = - \frac{N^2}{16 ( 2 \pi )^2} \int_{u=\epsilon} \frac{d\omega d^3q}{(2\pi)^4}\ 
 \sqrt{-\gamma}\ \frac{1}{2} (\ln(\epsilon) + \tilde{c}) F_{ij} F^{ij}\ ,
\end{equation}
where $\gamma$ represents the determinant of the metric induced by the bulk metric on the $u=\epsilon$ shell, the indexes $i,j$ run only over the 4
boundary directions and $\tilde{c}$ is a real numerical constant (we will later comment on this constant).
Notice that the overall factor is chosen to express $S_{\text{c.t.}}$ in terms of the gothic variables $\mathfrak{w}$ and $\mathfrak{q}$.
The renormalized action is obtained from
\begin{equation}
 S_{\text{ren}} = \lim_{\epsilon \rightarrow 0} \left[ S_{\text{reg}} + S_{\text{c.t.}}\right]\ ,
\end{equation}
and, explicitly, we have
\begin{equation}\label{onshelRen}
 S_{\text{ren}} =  \frac{(N T)^2}{16} \int \frac{d\omega d^3 q}{(2\pi)^4}\ 
 \phi_0 \left[ \phi_1 - {\phi}_0 (\mathfrak{w}^2 - \mathfrak{q}^2) \frac{c}{2} \right]\ ,
\end{equation}
where we have introduced $c = 2(1 - \tilde{c})$ for later convenience.

We remind the reader that $\phi_0$ represents the boundary value of the 
bulk fluctuation field which, through the gauge/gravity dictionary, is mapped to a source 
of the dual boundary theory. It is therefore with respect to $\phi_0$ that we are interested in taking 
derivatives of the on-shell bulk action. 
Specifically, the renormalized 2-point correlation function for the transverse current is obtained taking the second order 
derivative of the renormalized on-shell action (\ref{onshelRen}) with respect to $\phi_0$.
In Fourier space we obtain
\begin{equation}\label{dd}
 \begin{split}
 \frac{\delta^2 S_{\text{ren}}}{(\delta \phi_0)^2} 
 &=  \frac{(N T)^2}{16}\  \left[ \frac{\delta\phi_1}{\delta \phi_0} 
 - c(\mathfrak{w}^2 - \mathfrak{q}^2) \right]\\
 &\sim  \frac{(N T)^2}{16}\  \left[ \frac{\phi_1}{\phi_0} 
 - c(\mathfrak{w}^2 - \mathfrak{q}^2)  \right]\ .
 \end{split}
\end{equation}
In the last passage we exploited the linearity assumption and the fact 
that for zero source $\phi_0$ the fluctuation profile becomes trivial%
\footnote{More precisely, the differential problem \eqref{transverse},
being linear, is invariant under a rescaling of the field $\phi(u)$. 
The solution then depends actually on only one parameter; said otherwise,
$\phi_1$ is proportional to $\phi_0$.}.

From \eqref{dd} we can define the renormalized retarded correlator for the transverse current:
\begin{equation}\label{grin}
 G^{(c)}(\mathfrak{w},\mathfrak{q}) = -  \frac{(N T)^2}{16}\  \left[ \frac{\phi_1}{\phi_0} 
 - c(\mathfrak{w}^2 - \mathfrak{q}^2)  \right]\ ;
\end{equation}
the label $(c)$ is a reminder of the dependence of the correlation function on the coefficient $c$
in front of the contact term.
The correlation function \eqref{grin} is the fundamental quantity that allows us to compute the 
electromagnetic response of the strongly coupled plasma according to equations (\ref{eps_tra}) 
and \eqref{eps_traK}, as explained in Section \ref{scene}.
It is easy to check that the function defined in equation \eqref{grin} satisfies the usual properties for a response
function of a causal quantum field theory: namely it has poles in the negative imaginary part of the complex plane
and its imaginary part is negatively defined when $\omega$ and $q$ are real.

\subsubsection{Contact term}
\label{ct}

As explained above, the computation of the current-current correlator requires 
the renormalization of the on-shell action. 
In particular, the renormalization demands to 
consider a counter-term to cancel the logarithmic divergence in \eqref{asy_bou}. 
However, as usual in quantum field theory, the choice of the counter-term is not unique, and only  
its diverging part is specified by the renormalization procedure; the finite part should instead be 
fixed by some experimental measure. 
Such ambiguity is accounted by the real constant $c$ in the definition of the correlator in (\ref{grin}), that is 
actually not fixed by any consistency requirement related to symmetries and Ward identities.
The arbitrariness of the finite part of the counter-term 
introduces a term proportional to $\mathfrak{w}^2-\mathfrak{q}^2$ in the correlator (\ref{grin}) which is usually referred to as a \emph{contact term}%
\footnote{It is indeed a polynomial function of
$\mathfrak{w}$ and $\mathfrak{q}$ that corresponds to space-time delta-functions 
or derivatives thereof. These contributions are clearly
associated to the behavior of the correlator at coincident points.}.
Therefore, the constant $c$ in the correlator should be in principle fixed 
with an experimental measure of an observable containing it. 
Once the value of $c$ is fixed, the model does provide quantitative results depending on $N$ (a quantity
which, roughly speaking, is associated to the number of degrees of freedom of the system) and the temperature $T$ 
(which represents a physical scale of the model considered on the black hole solution). 
As we will see, it is particularly important to underline that the qualitative behavior of the electromagnetic modes 
in the medium, and more specifically of the refractive index, is a robust feature with respect to
the actual value of $c$.

\subsection{Electromagnetic modes}
\label{lw}

As explained in Section \ref{scene}, the possible electromagnetic transverse modes 
supported by the plasma are obtained solving the dispersion equation%
\footnote{This is actually the wave equation for the transverse component of the electric field inside the medium.} 
involving $\mathfrak{q}$ and $\mathfrak{w}$: 
\begin{equation}\label{lwe}
 \frac{\mathfrak{q}^2}{\mathfrak{w}^2} = 1 - \frac{4\pi e^2}{(2\pi T)^2 \mathfrak{w}^2}\; G^{(c)}(\mathfrak{w},\mathfrak{q})\ ,
\end{equation}
where $e$ is the electromagnetic coupling constant as previously introduced in (\ref{eps_tra})%
\footnote{In the gauge/gravity correspondence setting, $J$ is the conserved current of a global $U(1)$ symmetry
of the quantum field theory. In order to obtain the electromagnetic response of the medium, the standard procedure is to introduce 
in the QFT action the interaction term $e J \mathcal{A}$ 
where $\mathcal{A}$ is the electromagnetic field considered as \emph{external} and $e$ is the associated electromagnetic coupling 
constant considered to be perturbatively small.
The retarded correlator of the global current provides, at leading order in $e$, the exact result for the retarded correlator of the local current.}.
Substituting the explicit expression of the correlation function obtained in \eqref{grin}, the wave equation becomes
\begin{equation}\label{lwe2}
 \frac{\mathfrak{q}^2}{\mathfrak{w}^2} = 1 + \frac{1}{\mathfrak{w}^2} \
   \frac{e^2 N^2}{16\pi}\  \left[ \frac{\phi_1}{\phi_0} 
 - c(\mathfrak{w}^2 - \mathfrak{q}^2)  \right]\\ .
\end{equation}
Note that, since our system is conformally invariant at zero temperature, as long as we work with  
the dimensionless quantities $\mathfrak{w}$ and $\mathfrak{q}$,
we have no explicit dependence on $T$ in the wave equation. $T$ actually provides only a scale with respect to
which we measure the actual physical quantities. Furthermore it could be interesting to notice that 
in \eqref{lwe} the presence of a finite value of $c$ can be reabsorbed by an overall rescaling of the 
correlation function; more precisely
\begin{equation}\label{lwe3}
 G^{(c)}(\mathfrak{w},\mathfrak{q}) \leftrightarrow \frac{G^{(c=0)}(\mathfrak{w},\mathfrak{q})}{1 - \frac{c\, e^2 N^2}{16 \pi}} \ .
\end{equation}
The rescaling \eqref{lwe3} suggests that, as far as the study of the wave equation \eqref{lwe} is concerned,
it is possible to trade the contact term with a normalization factor 
in front of the correlator, this latter being related to $e N$.
We could therefore work without specifying the value of $e N$ and reducing everything to a contact term to be 
fixed against a physical measure performed on the electromagnetic modes.
One could for example think to fix $c$ with the requirement that the value of $\mathfrak{q}(\mathfrak{w}=0)$ 
for a specified mode coincides with that measured in an experiment.

To check the soundness of our results we did a scan of the electromagnetic modes over a broad range of values of $c$ and $e N$. 
The results show that the qualitative behaviour presented in Section \ref{results} is the same for a very broad range of parameters inside the validity of the numerical computation. 
For convenience in the paper we decided to plot the results for the specific values: $e N=3$ and $c=5.5$.

\section{Semi-numerical analysis and checks}
\label{nume}

In this Section we would like to explain  briefly how we actually performed the computations
and characterized the various electromagnetic modes supported by the strongly coupled plasma.
We refer again to Figures \ref{q}, \ref{n} and \ref{prop} in Section \ref{results}.
The aim is to solve the wave equation (\ref{lwe}) and find the dispersion relations 
$\mathfrak{q}_A=\mathfrak{q}_A(\mathfrak{w})$ (the label $A$ distinguishes
the different modes) connecting the complex wave-vector $\mathfrak{q}$ and the real 
frequency $\mathfrak{w}$.

At first we solve the equation for the transverse vector fluctuations \eqref{transverse}; 
once a solution is obtained, we can read the near-boundary coefficients and
plug them into (\ref{grin}) to find the correlator.
To solve the differential equation \eqref{transverse} 
we use two different methods whose results are later compared and cross-checked.
The first method consists in integrating numerically the equation of motion \eqref{transverse}
as explained in \cite{Hartnoll:2009sz}. The second method is semi-analytical and consists 
in expanding the function $\phi(u)$ near the horizon (as in \eqref{IR_exp}) and solving Equation \eqref{transverse} order 
by order along the lines of \cite{Kovtun:2005ev}; this infrared solution is then matched with an analogous term-wise 
solution computed at the boundary, see \eqref{UV_phi}.
It turns out that, while the numerical method yields more precise answers, the semi-analytical method is instead more agile.

Once a solution to \eqref{transverse} is obtained, it is then possible 
to follow the gauge/gravity recipe and define a correlation function as described in Subsection \ref{Green}.
For very small values of $\mathfrak{w}$ and $\mathfrak{q}$ we check that the correlation function agrees with the
analytical result of \cite{Policastro:2002se}. 
The correlator is then inserted in the electromagnetic wave equation (\ref{grin})
to search for the dispersion relations  $\mathfrak{q}_A=\mathfrak{q}_A(\mathfrak{w})$ of the various transverse modes.
As the correlator is a complicated rational function of $\mathfrak{q}$ and $\mathfrak{w}$, the electromagnetic wave equation 
has in general many solutions. 
The set of possible solutions $\mathfrak{q}_A(\mathfrak{w})$ can be 
studied relying on a simplifying approximation: We build the correlator using the previously introduced semi-analytical 
matched solution of \eqref{transverse}; we then write the correlator as 
the ratio of two polynomials in $\mathfrak{w}$ and $\mathfrak{q}$; we expand both 
the numerator and the denominator up to a suitable order in $\mathfrak{q}$ and $\mathfrak{w}$ around a chosen point in 
the $(\mathfrak{w},\mathfrak{q})$ space (this statements will be shortly made more precise).
The results we obtained with the aforementioned approximations are afterwards checked against the 
correlation function computed from the full numerical solution.

Let us be more specific on the computational procedure.
We find solutions to the electromagnetic wave equation expressed as complex functions $\mathfrak{q}_A(\mathfrak{w})$
where the frequency is a real quantity. We test the soundness of the solutions as follows:
we first repeat the procedure with an increased order in $\mathfrak{w}$ and $\mathfrak{q}$ up to which we expand the 
numerator and the denominator of the correlation 
function; we than check that the solutions found previously (i.e. with shallower expansions) still solve the equation 
within the numerical tolerance.
As an additional test we check that 
the solutions fall in the $(\mathfrak{w},\mathfrak{q})$ region 
where the approximated correlation function adhere well to the correlation function obtained 
in the fully numerical approach.
Eventually and more stringently we also check that the solutions obtained with the approximate method solve 
the electromagnetic wave equation written in terms of the fully numerical correlation function within the numerical tolerance.

We start looking for reliable electromagnetic wave solutions in the small $\mathfrak{w}$ and $\mathfrak{q}$ 
region ($\mathfrak{w},\mathfrak{q}\lesssim 1$); we then follow the modes iterating 
the approximate procedure and expanding around a generic point of their dispersion relation $\mathfrak{q}(\mathfrak{w})$. 
This allows us to analyze the modes at higher values of the frequency.

\section{Phenomenology: wavelengths and the QGP}
\label{QGP}

The analysis presented so far should be able to capture the electromagnetic response 
of a generic globally neutral strongly coupled plasma described by the gauge/gravity correspondence. 
A natural candidate for phenomenological application of the model at hand is  the QGP studied in high energy physics
experiments such as RHIC or LHC, where the plasma is dominated by the temperature, while the charge density is comparatively smaller.
A similar phenomenological investigation has been already performed in \cite{Noi2} for the hydrodynamical modes 
of strongly coupled and charged plasmas. However in \cite{Noi2} it was observed that, even if the fluid dynamic analysis 
demonstrates the generic presence of negative refraction, the wavelength of the electromagnetic mode
is larger than the dimension of the typical QGP sample produced in actual experiments.

In this Section, without pretending to provide any specific experimental or phenomenological prediction, 
we briefly investigate the wavelengths of the modes studied in the paper.
In particular we show that the wavelengths of some non-hydrodynamical modes 
are smaller than the those of the hydrodynamical modes considered in \cite{Noi2} and that they are 
actually comparable with the typical 
dimension of the sample (namely few fm). 
This result supports the possibility that the coupling of the electromagnetic field with 
the non-hydrodynamical quasi-normal modes  could in principle have phenomenological and experimental relevance and it calls for further investigation.

\begin{figure}[t]
\centering
\includegraphics[width=78mm]{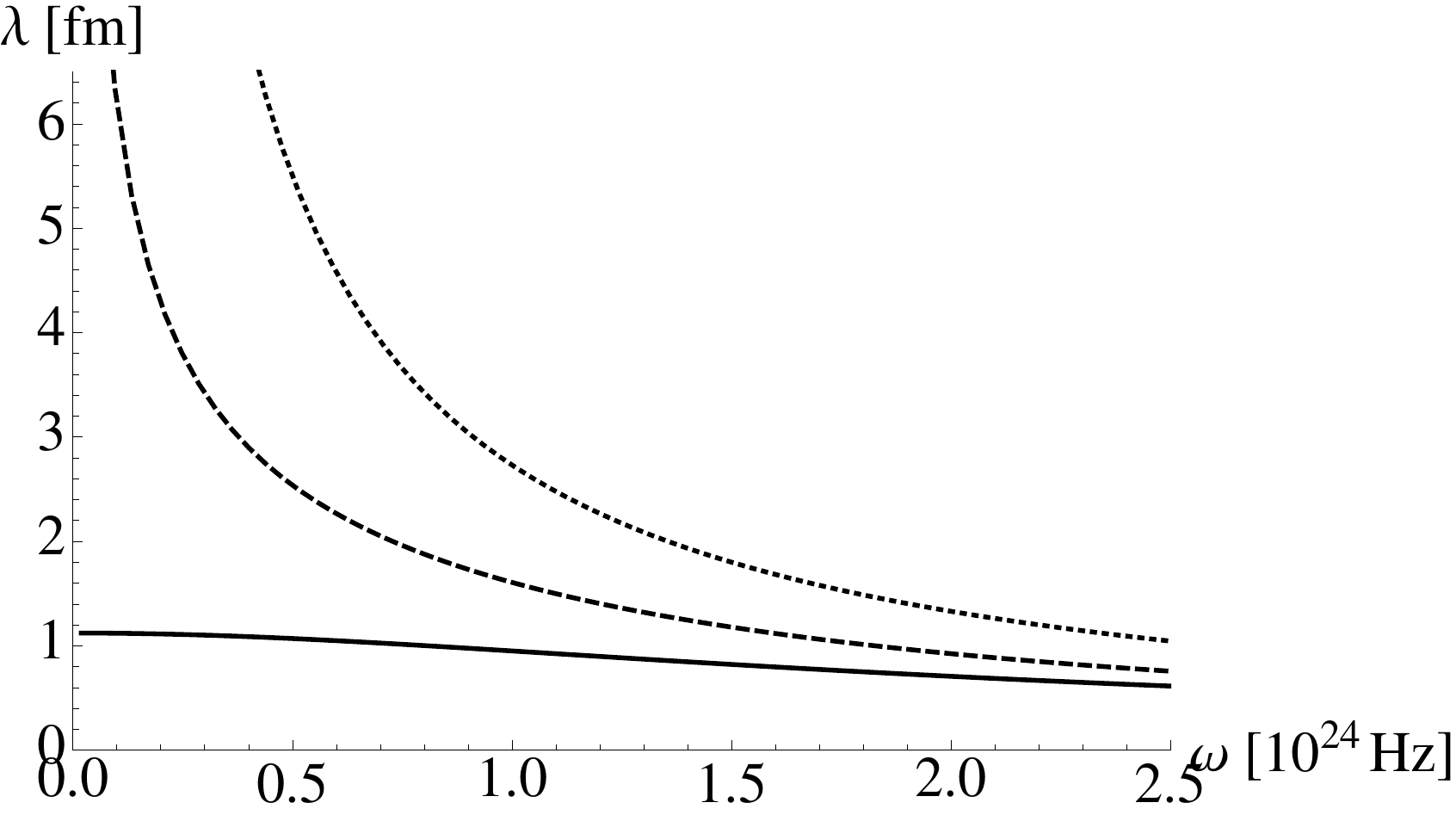} 
\caption{
Wavelengths of the transverse electromagnetic modes.
This values has been obtained considering $T=200$ MeV. We set $e N=3$ and $c=5.5$.}
\label{Olambda}
\end{figure}

In Figure \ref{Olambda} we report the wavelengths of our electromagnetic transverse modes. 
The wavelength of a mode is given by $\lambda = 2\pi/ \text{Re}[q] = 1/ T \text{Re}[\mathfrak{q}]$ in accordance with \eqref{Twq}.
To associate a physical order of magnitude to the wavelength of the analyzed modes we need 
to specify the temperature of interest. 
In relation to the QGP a typical value for the temperature is $200$ MeV for experiments like RIHC or LHC%
\footnote{Due to our numerical procedure, with the choice of temperature $T=200$ MeV, the regime of validity of our results 
corresponds to a frequency interval roughly between $10^{14}$ Hz and $10^{24}$ Hz.}.
Re-introducing the dimensionful physical constants, we find the following estimate for the wavelength
\begin{equation}
 \lambda = \frac{\hbar c}{T\; \text{Re}[\mathfrak{q}]} \sim 1\, \text{fm}\ ,
\end{equation}
for frequency of the order of $10^{24}$ Hz, as reported in Figure \ref{Olambda}.
In particular, note that the negative refracting mode of Figure \ref{Olambda} (solid line) presents a wavelength of order $1$ fm for the full range of frequency.

The size of the QGP samples in current experiments is actually of the order of few fm's \cite{CasalderreySolana:2011us,vanHees:2011vb,Aamodt:2011mr}, 
therefore (higher) non-hydrodynamical modes could actually probe the plasma and perhaps leave there their footprint.
We leave a deeper exploration of the higher modes and their possible effects on the QGP for future work.

\section{Longitudinal channel}
\label{long}

\begin{figure}[t]
\centering
\includegraphics[width=78mm]{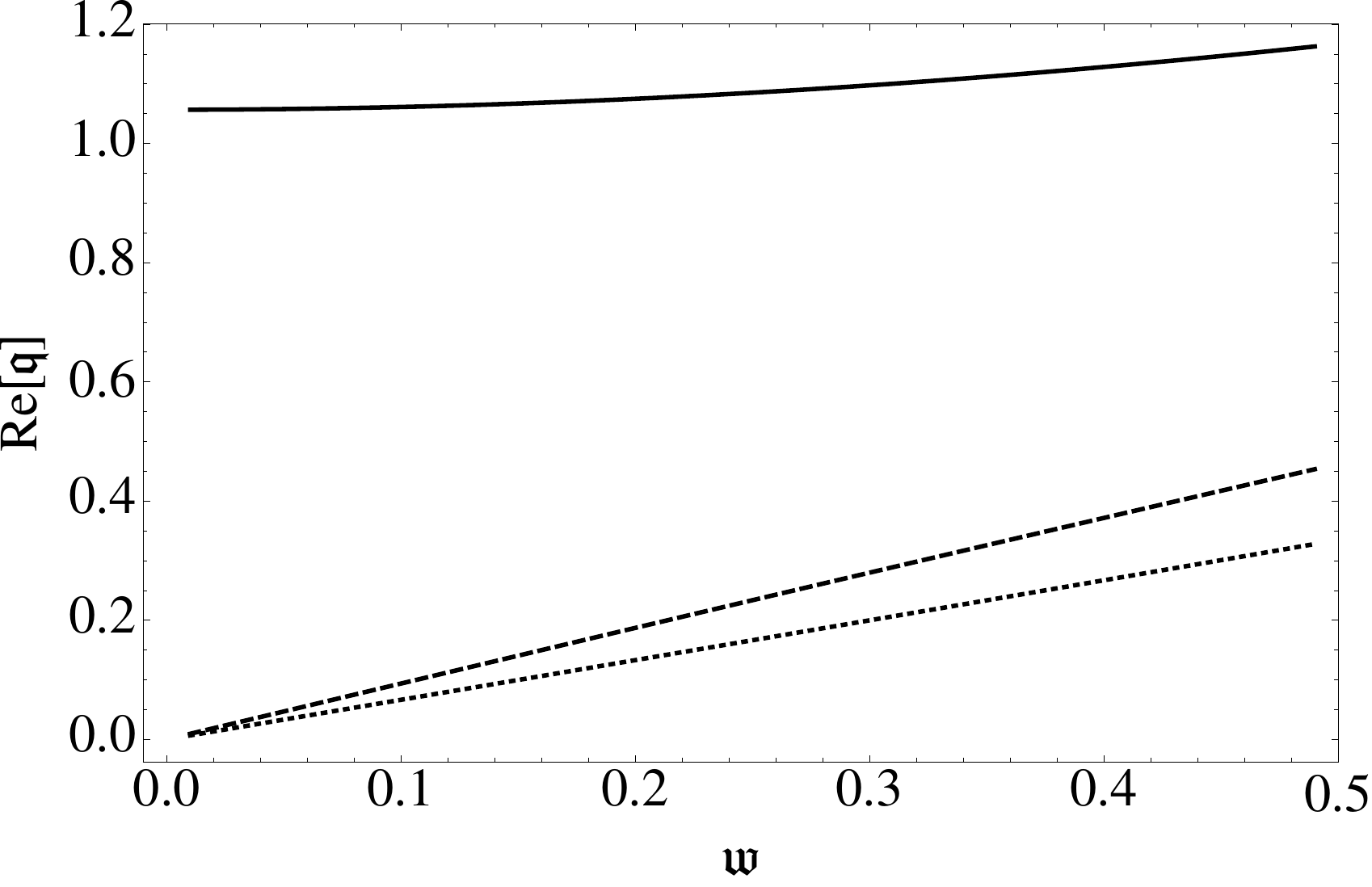} 
\includegraphics[width=78mm]{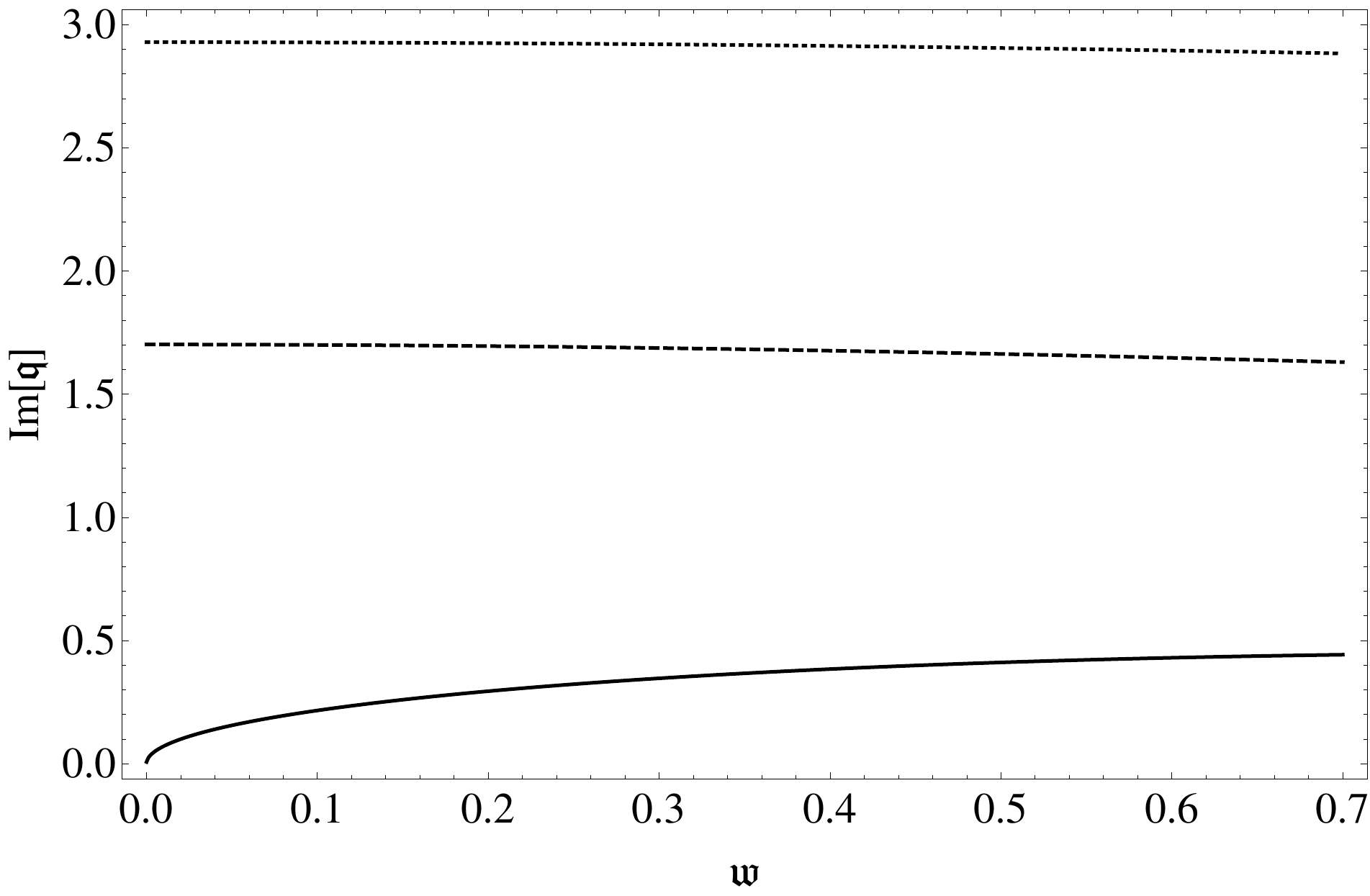} 
\caption{Real (left) and imaginary (right) parts of the rescaled wave-vector ${\mathfrak q}$ 
as a function of the rescaled frequency ${\mathfrak w}$ for the first three longitudinal electromagnetic modes.}
\label{longlight}
\end{figure}

In this Section we want to analyze briefly the longitudinal electromagnetic modes supported by the strongly coupled plasma
and we proceed with a similar spirit as with the transverse modes.
As we have previously recalled (see Section \ref{scene}), in the presence of spatial dispersion the response function of a medium 
depends also on the wave-number vector $q$. This fact implies that
also the longitudinal component of the electric field can propagate when the Maxwell equation for the longitudinal modes 
of the electric field (in Fourier space)
\begin{equation}\label{eps_lon}
 \epsilon_{L}(\omega,q) = 1 - 4 \pi e^2\, \frac{ {\cal G}_{L}(\omega,q)}{\omega^2}\ =0
\end{equation}
is satisfied. ${\cal G}_{L}(\omega,q)$ is the retarded correlator of the longitudinal current and, as in the 
transverse modes case, it is a rational function whose poles correspond to the longitudinal quasi-normal modes of the plasma. 
Equation (\ref{eps_lon}) provides a set of dispersion relations $q_A=q_A(\omega)$ between the complex longitudinal wave-vectors $q$ 
and the real frequency $\omega$ of the mode.
In this Section we simply give an account of some of the results whose computational details can be found
in Appendix \ref{lon_det}. 
In Figure \ref{longlight} and \ref{long_ref} we plot the real and imaginary parts of the wave-vector%
\footnote{As for the transverse modes, we use the rescaled quantities ${\mathfrak q}= q / 2 \pi T$,  ${\mathfrak w }= \omega / 2 \pi T$.}
$\mathfrak{q}$, the real part of the longitudinal refractive index $n=\mathfrak{q}/\mathfrak{w}$ and the ratio between the imaginary and the real parts of $n$
for the first three electromagnetic modes supported by the strongly coupled plasma. 
These Figures should be compared with Figures \ref{q}, \ref{n} and \ref{prop} in Section \ref{results} which instead refer to the transverse sector.

It is very interesting to observe that also the longitudinal channel supports various electromagnetic modes. 
However, we did not find any negative refractive longitudinal mode. 
All the modes have finite momentum at vanishing frequency. Nevertheless for two of them $\mathfrak{q}(\mathfrak{w}=0)$ is purely imaginary, 
while for the remaining mode is purely real. The modes having an imaginary momentum 
at vanishing frequency are highly dissipative in the IR regime even though the increase 
of the real part of $\mathfrak{q}$ with the frequency indicates that they could become more propagating 
at higher frequencies, as it can be seen in Figure \ref{long_ref}.
The mode with purely real momentum at $\mathfrak{w}=0$ is highly propagating 
already in the low-frequency region and it keeps this characteristic in the whole 
range of frequency that we considered. Notice that the sign 
of the real part of the momentum shows that all these 
modes have positive refracting index.

\begin{figure}[t]
\centering
\includegraphics[width=78mm]{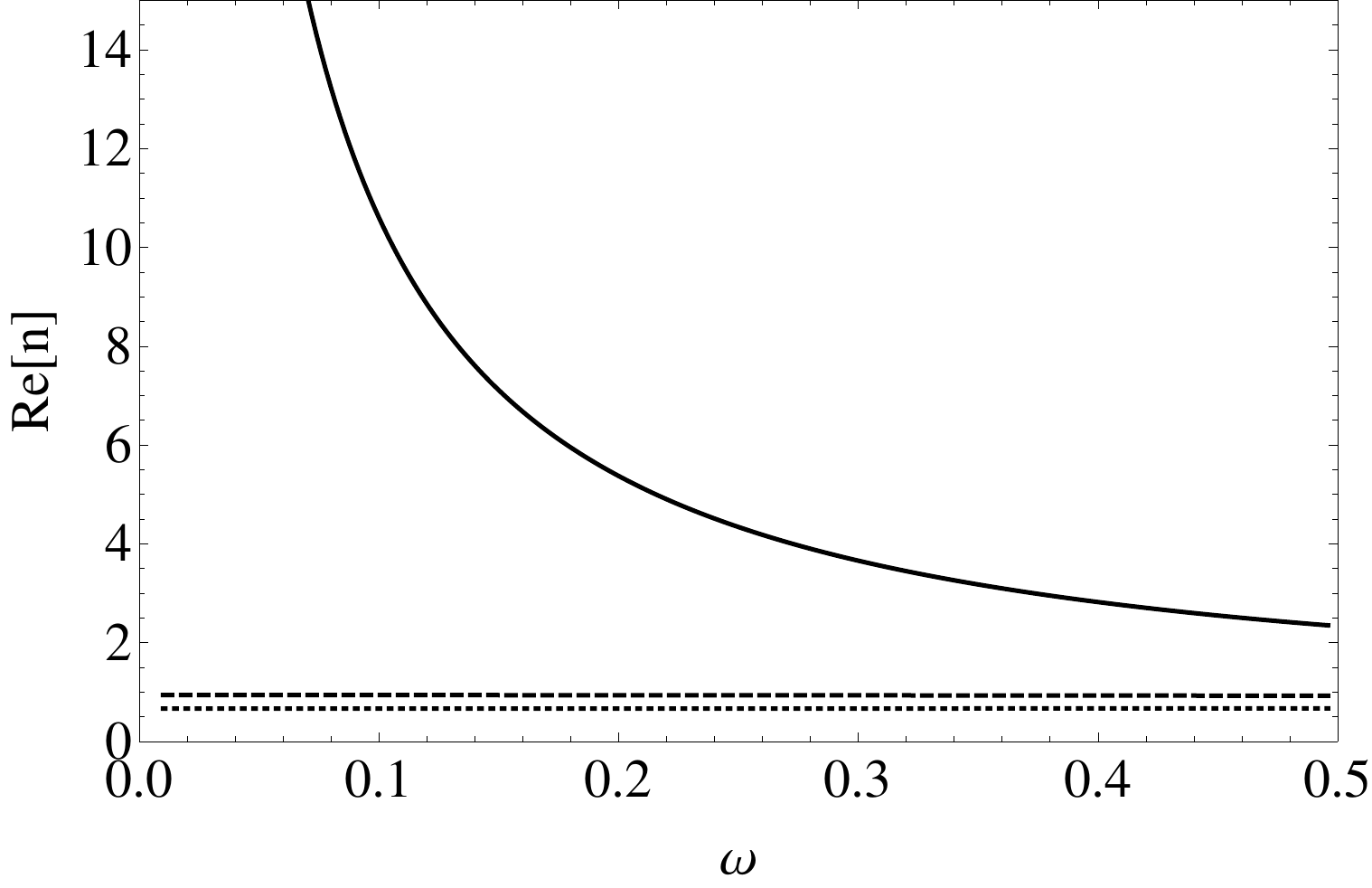} 
\includegraphics[width=78mm]{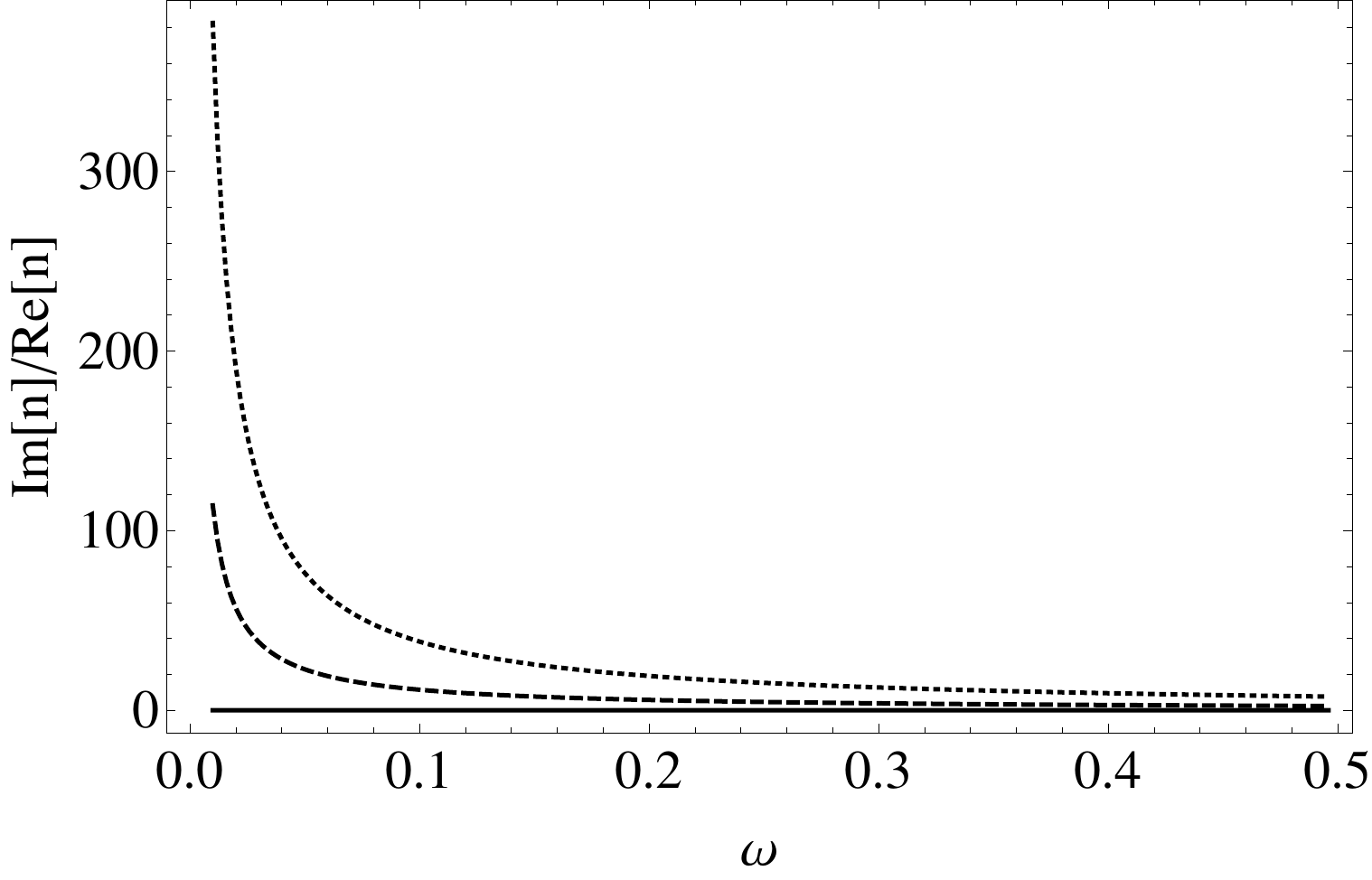} 
\caption{Real part of the refraction index $n$ and of the ratio $\text{Im}(n)/\text{Re}(n)$
as functions of the rescaled frequency ${\mathfrak w}$ for the first three longitudinal electromagnetic modes.}
\label{long_ref}
\end{figure}

\section{Conclusion}
\label{con}

In this paper we studied the electromagnetic linear response of strongly coupled 
neutral plasmas described by the gauge/gravity correspondence and characterized
the electromagnetic modes with the lowest wave-vectors. 
The salient features of the present analysis are the presence of multiple electromagnetic 
waves with different refractive indexes and a propagating negative refracting mode with 
very small dissipation. Our study has been performed without adopting any hydrodynamical 
approximation; hence we extended some previous hydrodynamical studies of strongly coupled 
plasmas beyond the regime of small frequencies and wave-vectors.

Our simple model highlights the potential richness of the electrodynamic response of strongly coupled plasmas and this calls for further investigations. 
In particular, on a phenomenological level, we showed that the characteristic wavelengths of the electromagnetic modes in the plasma 
could be comparable to the typical size of the QGP samples produced in high-energy physics experiments. 
This fact provides some arguments supporting the possible relevance of the presented exotic phenomena for actual physical systems.

Our results suggest many future lines of investigation. 
Similar systematic analyses of the electromagnetic properties can be indeed performed using different kinds of 
gravitational backgrounds featuring appealing characteristics such as finite charge density, 
spontaneous symmetry breaking, non-relativistic or non-isotropic setups and the presence of magnetic fields. 
These extensions could find application in various physical systems such as the QGP but also in condensed matter
and astrophysics.

We know that strong spatial dispersion is a crucial ingredient in producing the exotic electromagnetic response
we are concerned with, however it would be desirable to understand whether some more precise connection between the modes of the plasma and its
associated electromagnetic modes can be clarified. 
In particular it would be nice to have a direct understanding of the presence or not of negative refraction for the various electromagnetic modes
based on the QNM structure.

An interesting and ambitious aim would be to provide a systematic and complete analysis of all the non-hydrodynamical modes.

\section*{Acknowledgments}

It is our great pleasure to acknowledge Alberto Mariotti for relevant discussions in the first phases of the project and Antonio Amariti for very useful suggestions and for precious comments on the draft.
Moreover we would like to thanks Carlo Maria Becchi,  Riccardo Argurio,  Giovanni Villadoro,
 Diego Redigolo, Flavio Porri, Matteo Bertolini,
 Andrea Amoretti,  Nicola Maggiore,  Nicodemo Magnoli,
 Alessandro Braggio, Antonello Scardicchio, Sebastiano Pilati, 
Grazia Luparello, Luciano Ramello, Shira Chapman for interesting discussions and for having shared with us their valuable insight. D.F would like to acknowledge the kind hospitality of the LPTHE, where part of this research has been implemented.

D.F. is "Charg\'e de recherches" of the Fonds de la Recherche Scientifique F.R.S.-FNRS (Belgium), and his research is supported by the F.R.S.-FNRS.

The work of D.F and A.M. is partially supported by IISN - Belgium
(conventions 4.4511.06 and 4.4514.08), by the \textquotedblleft Communaut%
\'{e} Fran\c{c}aise de Belgique" through the ARC program and by the ERC
through the \textquotedblleft SyDuGraM" Advanced Grant.

\appendix

\section{Details about the longitudinal channel}
\label{lon_det}

In this appendix we would like to provide some details on the actual computation done for the longitudinal electromagnetic modes. 
The procedure is similar to that explained in the main text for the transverse modes. However there exist some differences to be taken into account.
Using the rotation invariance of the system we fix the four momentum as 
$k=(\omega,0,0,q)$. The longitudinal current then corresponds to the $z$ direction of the field $A$ in the dual gravity setup: $A_z$. 
However in this case $A_z$ mixes with the time component $A_t$ \cite{Kovtun:2005ev}.
We consider the following ansatz
\begin{equation}
 A_m dx^m = A_t e^{-i(\omega t - q z)} dt + A_z e^{-i(\omega t - q z)} dz\ .
\end{equation}
and define the gauge invariant combination:
\begin{equation}\label{psi_def}
 \psi = \mathfrak{q} A_t + \mathfrak{w} A_z\ ,
\end{equation}
which represents the electric field in the $z$ direction. 
The $A_t$ and $A_z$ equations then leads to a single equation for $\psi$, namely
\begin{equation}
 \psi''(u) + \frac{\mathfrak{w}^2 f'(u)}{f(u)\left[\mathfrak{w}^2 - \mathfrak{q}^2 f(u)\right]} \psi'(u) 
 + \frac{\mathfrak{w}^2 - \mathfrak{q}^2 f(u)}{u f(u)^2} \psi(u) = 0\ .
\end{equation}
From a near-boundary study we find the asymptotic behavior 
of the field
\begin{equation}
 \psi(u) = \psi_0 + u \psi_1 + u \ln(u) \tilde{\psi}_1 + u^2 \psi_2 + u^2 \ln(u) \tilde{\psi}_2 + ...\ ,
\end{equation}
which is identical to that of $\phi$ written in \eqref{UV_phi}. Recall indeed that in the limit of vanishing 
$q$ the longitudinal and the transverse sector coincide. Moving to finite $q$ 
does not affect the general form of the UV asymptotic expansions. 
The same UV asymptotic behavior for the longitudinal and traverse modes 
implies that the holographic renormalization procedure is analogous 
in both channels.We can indeed use the same counter-term 
$c(\mathfrak{w}^2 - \mathfrak{q}^2)$
 for both polarizations.

The on-shell action for the combination (\ref{psi_def}) is:
\begin{equation}\label{os_lag_lon}
  \frac{(N T)^2}{16}
  \int  \frac{d\omega d^3 q}{(2 \pi)^4}
  \frac{f(u)}{\mathfrak{q}^2 f(u) - \mathfrak{w}^2}\ \psi' \psi\ ;
\end{equation}
This is the longitudinal version of Equation \eqref{os_lag_tra}. Hence the  $A'_z A_z$ term of the on-shell 
action is 
\begin{equation}
 \frac{(N T)^2}{16}
 \int  \frac{d\omega d^3 q}{(2 \pi)^4}
 \frac{\mathfrak{w}^2 f(u)}{\mathfrak{q}^2 f(u) - \mathfrak{w}^2}\ A_z' A_z\ .
\end{equation}
We understand that, in a similar way as for the transverse channels, the $zz$ current-current correlator related to the second functional differentiation
of the on-shell action with respect to $A_z$ can be expressed as follows
\begin{equation}
 G_{zz}^{(c)}(\mathfrak{w},\mathfrak{q}) = - \frac{(N T)^2}{16} \frac{\mathfrak{w}^2}{\mathfrak{q}^2-\mathfrak{w}^2} 
 \left[\frac{\psi_1}{\psi_0} - c (\mathfrak{w}^2-\mathfrak{q}^2)\right]\ .
\end{equation}
These last passages are done in line with \cite{Kovtun:2005ev} to which we refer for further details.
Note that the $tt$ and $tz$ correlators are obtained in a similar way leading to
\begin{equation}
 G_{tt}^{(c)}(\mathfrak{w},\mathfrak{q}) = - \frac{(N T)^2}{16} \frac{\mathfrak{q}^2}{\mathfrak{q}^2-\mathfrak{w}^2} 
 \left[\frac{\psi_1}{\psi_0} - c (\mathfrak{w}^2-\mathfrak{q}^2)\right]
\end{equation}
and
\begin{equation}
 G_{tz}^{(c)}(\mathfrak{w},\mathfrak{q}) = - \frac{(N T)^2}{16} \frac{\mathfrak{w} \mathfrak{q}}{\mathfrak{q}^2-\mathfrak{w}^2} 
 \left[\frac{\psi_1}{\psi_0} - c (\mathfrak{w}^2-\mathfrak{q}^2)\right]\ .
\end{equation}

It is important to notice that the set of $zz$, $tt$ and $tz$ correlators satisfy
the Ward identity
\begin{equation}
 k^\mu G^{(c)}_{\mu\nu} (\mathfrak{w},\mathfrak{q}) = 0\ .
\end{equation}
This fact is related to the structure of the frequency and momentum dependent factors in front 
of the expressions of the correlators. As a consequence, the contact term proportional 
to $c$ is not fixed by the Ward identities or, in other words, it is not constrained by 
symmetry requirements. All the arguments about the contact terms that we have described in relation 
to the transverse sector can be repeated for the longitudinal sector.

\section{Quasi-normal modes}

In order to characterize better the strongly coupled plasma under study in the main text,
we report here an analysis of its internal modes.
The retarded correlation function accounting for the electromagnetic response of the plasma presents poles at specific values for the momentum and 
frequency of the external perturbation. 
Such poles correspond in the dual gravitational picture to quasi-normal modes of the black hole solution (see for instance \cite{Kokkotas:1999bd,Kovtun:2005ev,Nollert:1999ji,Horowitz:1999jd,Son:2002sd,Amado:2007yr,Amado:2008ji,Starinets:2002br,Nunez:2003eq}).
To find the dispersion relations between $\mathfrak{q}$ and $\mathfrak{w}$ of the above-mentioned quasi-normal modes it is enough to look at
the zeros of the inverse of the correlation function. 
These solutions can be represented
(for instance) as complex functions $\mathfrak{q}(\mathfrak{w})$ of the real frequency%
\footnote{It is important to underline that these dispersion relations between $\mathfrak{q}$ and $\mathfrak{w}$ are 
the ones associated to the {\it quasi-normal modes} of the plasma and they are different from the dispersion relations for the {\it electromagnetic modes} 
supported by the plasma and discussed in the main text. Indeed the first ones come from solving the 
equations: $(G_{T,L}(\omega,q))^{-1}=0$, while the second ones solve equations which look approximately as
$q^2 = \omega^2 - G_{T}(\omega,q)$ and $\omega^2 - G_{L}(\omega,q)=0$.}.
In this Appendix we perform a characterization of the first quasi-normal modes of the plasma both in the longitudinal and transverse sectors.

\begin{figure}[t!]
\centering
\includegraphics[width=78mm]{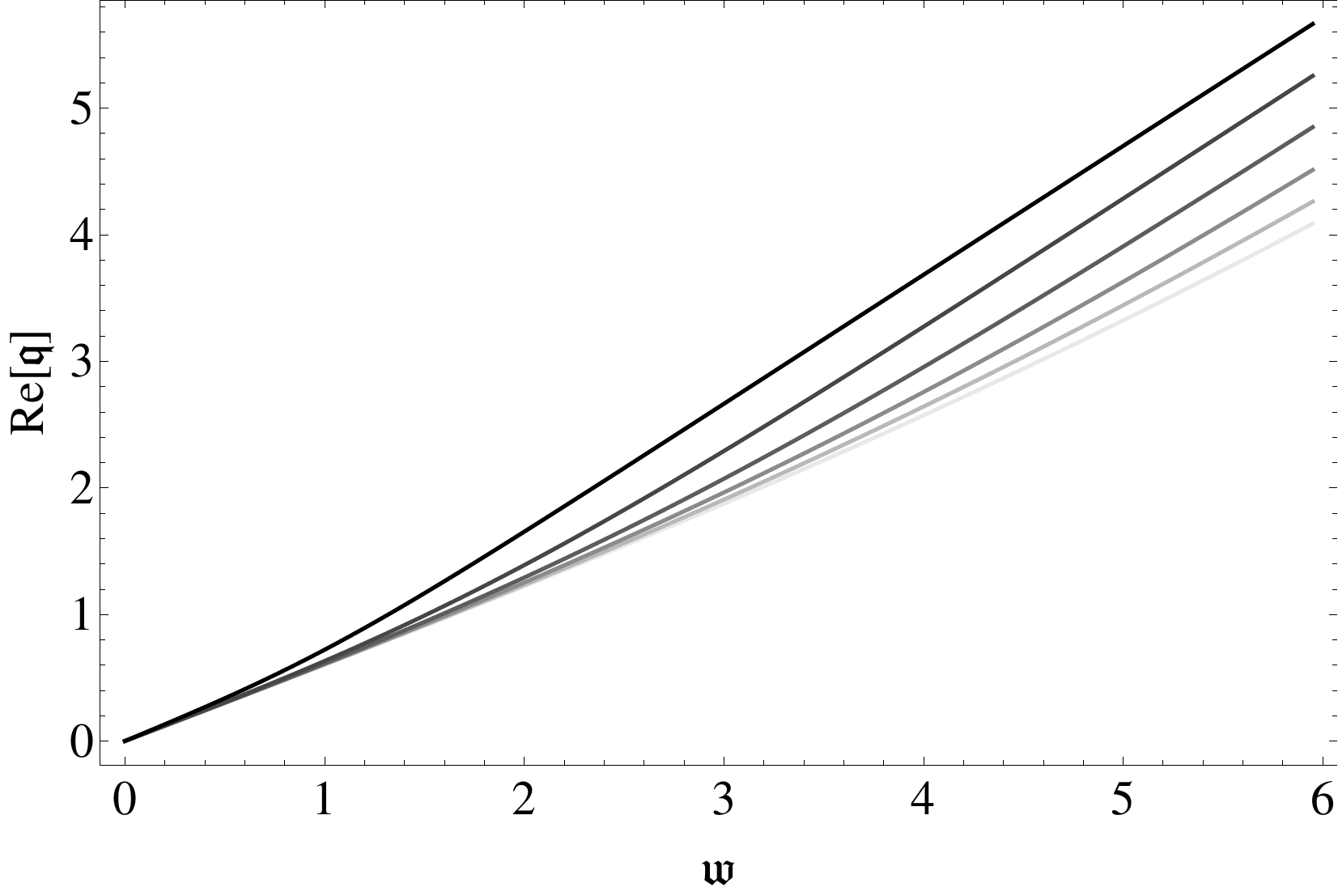} 
\includegraphics[width=78mm]{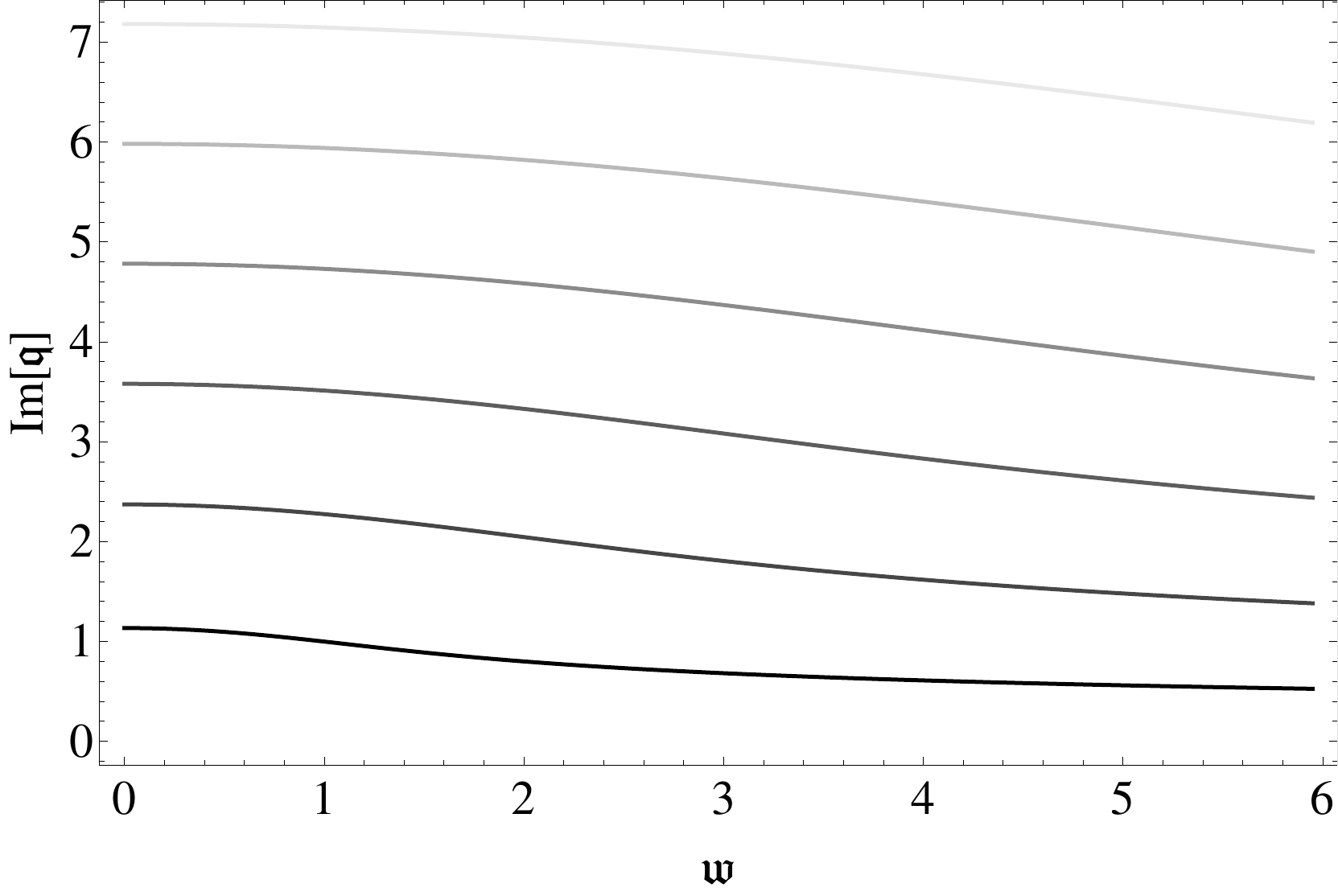} 
\caption{Real (left) and imaginary (right) parts of the rescaled wave-vector ${\mathfrak q}$ 
as a function of the rescaled frequency ${\mathfrak w}$ for the first transverse quasi-normal modes.
The numerical analysis suggest the presence of an infinite tower of analogous modes.}
\label{trans_mech}
\end{figure}

The results for the transverse sector are summarized in Figure \ref{trans_mech}. 
All the modes we found are qualitatively similar: they all show a finite imaginary part and 
null real part for $\mathfrak{q}$ as the frequency goes to zero. 
\begin{figure}[t!]
\centering
\includegraphics[width=78mm]{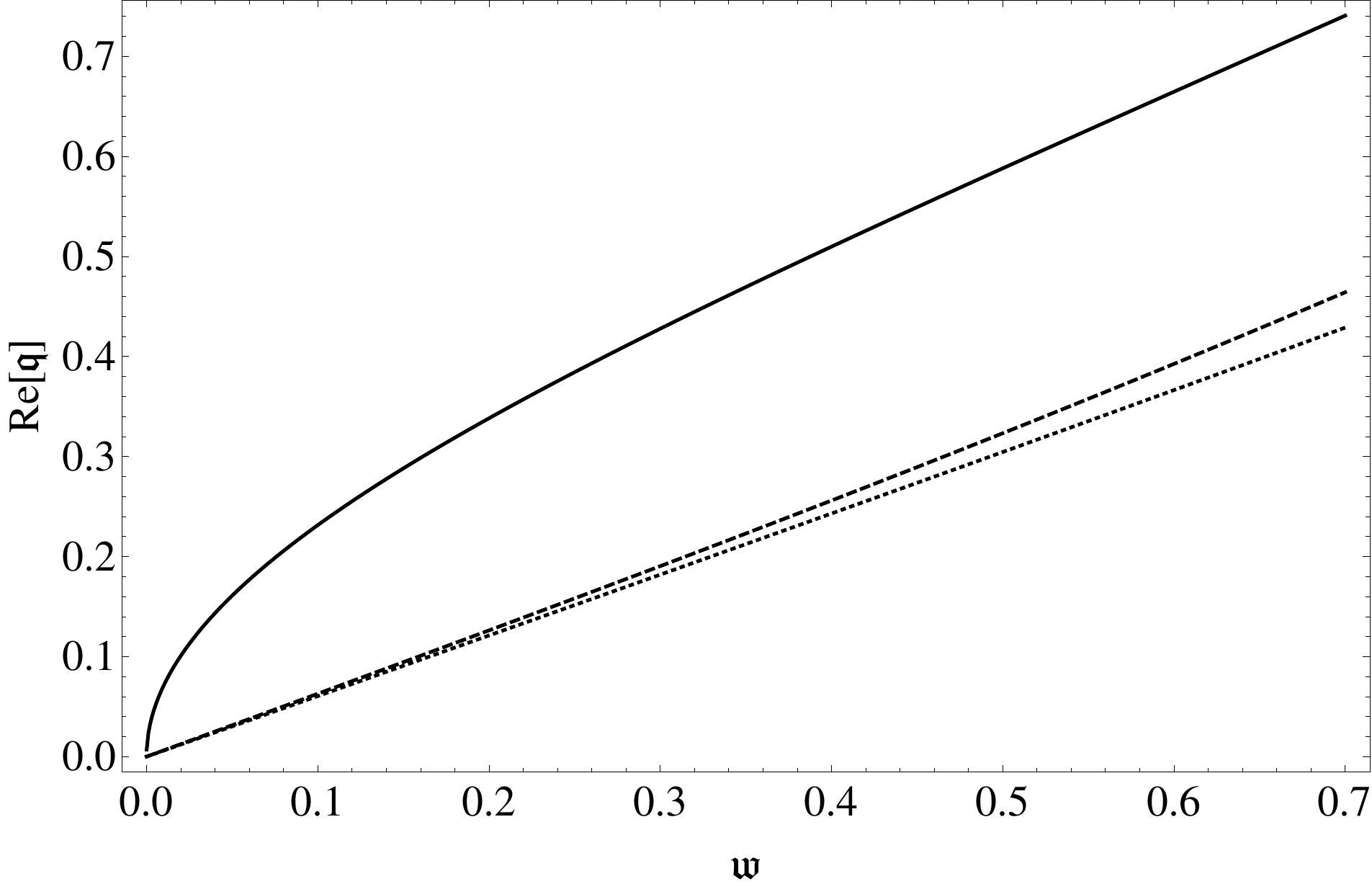} 
\includegraphics[width=78mm]{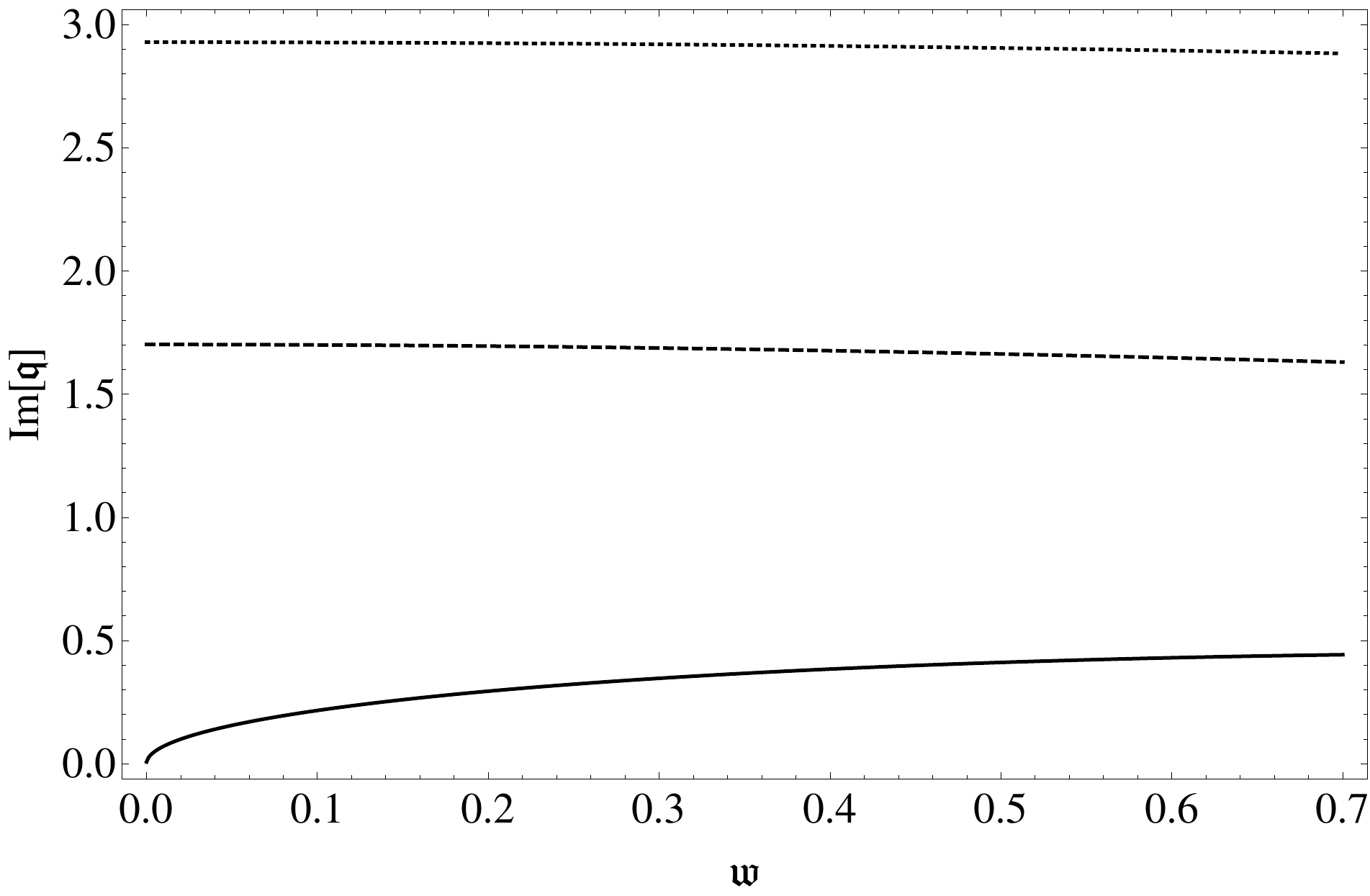} 
\caption{Real (left) and imaginary (right) parts of the rescaled wave-vector ${\mathfrak q}$ 
as a function of the rescaled frequency ${\mathfrak w}$ for the first three longitudinal quasi-normal modes.}
\label{long_mech}
\end{figure}
Such circumstance corresponds to having a highly dissipative (and non-propagating)
set of modes at low frequency. At higher values of the frequency, however, the real part of $\mathfrak{q}$
increases until it becomes significantly bigger than its imaginary part.
The modes are then propagating for higher values of the dimensionless frequency $\mathfrak{w}$.
 
The longitudinal sector features different kinds of modes (see Figure \ref{long_mech}): we have a diffusive 
mode whose complex momentum vanishes as the frequency goes to zero. 
This is the hydrodynamical longitudinal mode discussed%
\footnote{We also checked explicitly that at low frequency this longitudinal hydrodynamical mode
is well approximated by the diffusive pole structure with diffusion constant $D = 2\pi T$ 
(referred to dimensionful $\omega$ and $k$).}
in \cite{Policastro:2002se}.
The remaining modes are closely analogous to those we found in 
the transverse sector. They have a purely imaginary momentum $\mathfrak{q}$ as the frequency vanishes.
The numerical investigation leads us to imagine that we have an infinite tower of such modes which again are strongly dissipating at low frequency.

As a final comment we want to underline that whenever possible we checked our results with \cite{Amado:2008ji}.

\end{document}